\documentclass[12pt,fleqn]{article}
\pdfpageattr {/Group << /S /Transparency /I true /CS /DeviceRGB>>}

\usepackage[DIV12]{typearea}
\usepackage{amsmath,amsfonts,amssymb}
\usepackage{graphicx}
\usepackage{cite}
\usepackage{mathtools}
\usepackage{mathrsfs}
\usepackage{bbm}
\usepackage{pifont}
\usepackage{marvosym}
\usepackage{units}
\usepackage{cancel}
\usepackage[small,loose,md,TABBOTCAP]{subfigure}  
\usepackage{xcolor}
\usepackage{braket}
\usepackage[normalem]{ulem}

\usepackage{hyperref}

\newcommand{\Eqref}[1]{equation~\eqref{#1}}
\newcommand{\Figref}[1]{figure~\ref{#1}}
\newcommand{\Tabref}[1]{table~\ref{#1}}
\newcommand{\Secref}[1]{section~\ref{#1}}
\newcommand{\Appref}[1]{appendix~\ref{#1}}

\newcommand{\SemiDirect}[0]{\ensuremath{\rtimes}}

\newcommand{\eVdist}{\kern-0.06em}

\DeclareMathOperator{\im}{Im}
\DeclareMathOperator{\tr}{tr}
\DeclareMathOperator{\diag}{diag}

\DeclareMathOperator{\ord}{ord}

\newcommand{\D}{\mathrm{d}}
\newcommand{\I}{\mathrm{i}}
\newcommand{\ChargeC}{\mathcal{C}}

\newcommand{\ParityP}{\mathcal{P}}

\newcommand{\SU}[1]{\ensuremath{\mathrm{SU}(#1)}}

\newcommand{\U}[1]{\ensuremath{\mathrm{U}(#1)}}
\newcommand{\Z}[1]{\ensuremath{\mathbbm{Z}_{#1}}} 

\newcommand*{\rep}[2][]{\ensuremath{{\boldsymbol{#2}#1}}} 

\renewcommand{\bar}[1]{\overline{#1}}

\newcommand{\Tprime}{\ensuremath{\mathrm{T}'}}
\newcommand{\Afour}{\ensuremath{A_4}}

\newcommand{\DiscreteGroup}{\ensuremath{G}}
\newcommand{\UCP}{\ensuremath{U_\mathrm{CP}}}

\newcommand{\FSI}{\ensuremath{\mathrm{FS}_u}}
\newcommand{\TFS}[1]{\ensuremath{\mathrm{FS}_{#1}}}
\newcommand{\UU}[2][]{\ensuremath{U_{\rep[#1]{#2}}}}
\newcommand{\WW}[2][]{\ensuremath{W_{\rep[#1]{#2}}}}
\newcommand{\SIGMA}[2][]{\ensuremath{\Sigma_{\rep[#1]{#2}}}}
\newcommand{\rhoR}[1]{\ensuremath{\rho_{\rep[#1]{r}}}}
\newcommand{\chiR}[1]{\ensuremath{\chi_{\rep[#1]{r}}}}

\def\myvec{\empty}

\usepackage{pdfcomment}

\newcommand{\HLS}{HLS~\cite{Holthausen:2012dk}}
\newcommand{\CPgen}{\ensuremath{\boldsymbol{\widetilde{\ChargeC\ParityP}}}}

\allowdisplaybreaks[1]

\numberwithin{equation}{section}
\numberwithin{table}{section}

\def\mytitle{CP Violation from Finite Groups
}
\title{\mytitle}

\begin{document}

\begin{titlepage}

\begin{flushright}
 UCI-TR-2014-01\\
 TUM-HEP 929/14\\
 FLAVOR-EU-64\\
 CU-HEP-584\\
\end{flushright}

\vspace*{1.0cm}

\renewcommand*{\thefootnote}{\fnsymbol{footnote}}
\begin{center}
{\Large\textbf{\mytitle}}
\renewcommand*{\thefootnote}{\arabic{footnote}}

\vspace{1cm}

\textbf{Mu--Chun Chen
$^a$,}
\textbf{Maximilian Fallbacher
$^b$,} 
\textbf{K.T.~Mahanthappa
$^c$,}
\\
\textbf{
Michael Ratz
$^b$ and
Andreas Trautner
$^{b,d}$}
\\[3mm]
\textit{\small
{}$^a$ 
Department of Physics and Astronomy, University of California,\\
~~Irvine, California 92697--4575, USA
}
\\[3mm]
\textit{\small
{}$^b$ 
Physik Department T30, Technische Universit\"at M\"unchen, \\
~~James--Franck--Stra\ss e~1, 85748 Garching, Germany
}
\\[3mm]
\textit{\small
{}$^c$ 
Department of Physics, University of Colorado,\\
~~Boulder, Colorado 80309, USA
}
\\[3mm]
\textit{\small
{}$^d$ 
Excellence Cluster Universe, \\
Boltzmannstra\ss e~2, 85748 Garching, Germany
}
\end{center}

\vspace{1cm}

\begin{abstract}
We discuss the origin of CP violation in settings with a discrete (flavor)
symmetry \DiscreteGroup. We show that physical CP transformations always have to
be class--inverting automorphisms of \DiscreteGroup. This allows us to
categorize finite groups into three types:  (i) Groups that do not exhibit such
an automorphism and, therefore, in generic settings, explicitly violate CP. In
settings based on such groups, CP violation can have pure group--theoretic
origin and can be related to the complexity of some Clebsch--Gordan
coefficients. (ii) Groups for which one can find a CP basis in which all the
Clebsch--Gordan coefficients are real. For such groups, imposing CP invariance
restricts the phases of coupling coefficients. (iii) Groups that do not admit
real Clebsch--Gordan coefficients but possess a class--inverting  automorphism
that can be used to define a proper (generalized) CP transformation. For such
groups, imposing CP invariance can lead to an additional symmetry that forbids
certain couplings. We make use of the so--called twisted Frobenius--Schur
indicator to distinguish between the three types of discrete groups. With
$\Delta(27)$, \Tprime, and $\Sigma(72)$ we present one explicit example for each
type of group, thereby illustrating the CP properties of models based on them.
We also show that certain operations that have been dubbed generalized CP
transformations in the recent literature do not lead to physical CP
conservation. 
\end{abstract}
\end{titlepage}

\enlargethispage{0.5cm}
\section{Introduction and outline}

It is well known that the simultaneous action of parity and charge conjugation
(CP) is not a symmetry of Nature. This fact has been established experimentally
in oscillations and decays of $K$, $B$, and $D$ mesons.  Furthermore, CP
violation is a necessary condition to generate the observed matter--antimatter
asymmetry of the universe \cite{Sakharov:1967dj}. The origin of CP violation
is, thus, one of the most fundamental questions in particle physics.
Currently, all direct evidence for CP violation in Nature can be related to the
flavor structure of the standard model (SM) of particle physics
\cite{Kobayashi:1973fv}.

Since it is conceivable that the flavor structure may be explained by an
(explicitly or spontaneously broken) horizontal or flavor symmetry, it appears
natural to seek a connection between the fundamental origins of CP violation and
flavor. In the past, it has been argued that the appearance of complex
Clebsch--Gordan (CG) coefficients in some of these groups gives rise to
(explicit) CP violation \cite{Chen:2009gf}, thus relating CP violation to
some intrinsic properties of the flavor symmetry. 

There are many ways to check whether or not CP is (explicitly) violated in a
given setting. In the low--energy effective theory, an unambiguous check 
of the existence of CP violation is the computation of so--called weak basis invariants which, if vanishing, guarantee
the absence of (flavor related) CP violation
\cite{Bernabeu:1986fc,Gronau:1986xb,Branco:1986gr}.  However, in order to decide
whether CP violation is explicit or spontaneous in the high energy theory, one
has to identify the corresponding symmetry transformation that, if unbroken,
guarantees the absence of CP violation --- which is typically less
straightforward. Especially in settings with a discrete (flavor) symmetry, the
true physical CP transformation may be obscured by the fact that, say, 
complex Clebsch--Gordan coefficients are present.

To decide whether a particular transformation really conserves CP, one has to
check if it can be ``undone'' by a symmetry or basis transformation. If this is
the case, CP is conserved, otherwise CP is violated (cf.\ e.g.\
\cite{Lebedev:2002wq}). This then leads to the notion of a so--called
generalized CP transformation \cite{Ecker:1981wv,Ecker:1983hz,Branco:1999fs},
where one amends the canonical quantum field theory (QFT) transformation laws by
this operation.

The main purpose of this study is to explore the relation between discrete
(flavor) symmetries \DiscreteGroup\ and physical CP invariance guaranteed by
generalized CP transformations in more detail.

The outline of this paper is as follows. In \Secref{sec:generalizedCP} we
discuss the general properties of generalized CP transformations. In particular,
we will show that physical CP transformations are always connected to
class--inverting automorphisms of \DiscreteGroup. We will classify discrete
groups \DiscreteGroup\ based on the existence and the specific properties of
such transformations.  This will allow us to conclude that in theories based on
a certain type of symmetry CP is generically violated since one cannot define a
proper CP transformation. Section~\ref{sec:Examples} contains three examples
illustrating our results. In particular, we demonstrate that both explicit CP
violation and spontaneous CP violation with a phase predicted by group theory
can arise based on a decay example in an explicit toy model. As we shall see,
some of the transformations that were dubbed ``generalized CP transformations''
in the recent literature do not correspond to physical CP transformations.
Finally, in \Secref{sec:Conclusions} we summarize our results. In various
appendices we collect some of the more technical details relevant to our
discussion.

\section{Generalized CP transformations}
\label{sec:generalizedCP}

\subsection{The canonical CP transformation}

We start out by reviewing the standard transformation laws of quantum field
theory. By definition, charge conjugation
reverses the sign of conserved currents,
$j^\mu\xmapsto{\boldsymbol{\ChargeC}}-j^\mu$. For a scalar field operator
\begin{equation}
 \boldsymbol{\phi}(x)
 ~=~
 \int\!\D^3p\,\frac{1}{2E_{\vec p}}\,\left[
 \boldsymbol{a}(\vec p)\,\mathrm{e}^{-\I\,p\cdot x}
 +
 \boldsymbol{b}^\dagger(\vec p)\,\mathrm{e}^{\I\,p\cdot x}
 \right]\;, 
\end{equation}
this implies that the creation and annihilation operators $\boldsymbol{a}$ and
$\boldsymbol{b}$ for particles and anti--particles get exchanged. 
Combining this with a spatial inversion, i.e.\ a parity transformation,
the combined transformation is given by (e.g.\ \cite{Sozzi:2008zza})
\begin{subequations}\label{eq:fieldCP}
\begin{align}
 \left(\boldsymbol{\ChargeC}\,\boldsymbol{\ParityP}\right)^{-1}\,\boldsymbol{a}(\vec p)\,\boldsymbol{\ChargeC}
 \,\boldsymbol{\ParityP}
 &~=~\eta_{\ChargeC\ParityP}\,\boldsymbol{b}(-\vec p)
 \;, & 
 \left(\boldsymbol{\ChargeC}\,\boldsymbol{\ParityP}\right)^{-1}\,\boldsymbol{a}^\dagger(\vec p)\,\boldsymbol{\ChargeC}
 \,\boldsymbol{\ParityP}
 &~=~\eta_{\ChargeC\ParityP}^*\,\boldsymbol{b}^\dagger(-\vec p)
 \;,\\
 \left(\boldsymbol{\ChargeC}\,\boldsymbol{\ParityP}\right)^{-1}\,\boldsymbol{b}(\vec p)\,\boldsymbol{\ChargeC}
 \,\boldsymbol{\ParityP}
 &~=~\eta_{\ChargeC\ParityP}^*\,\boldsymbol{a}(-\vec p)
  \;, & 
 \left(\boldsymbol{\ChargeC}\,\boldsymbol{\ParityP}\right)^{-1}\,\boldsymbol{b}^\dagger(\vec p)\,\boldsymbol{\ChargeC}
 \,\boldsymbol{\ParityP}
 &~=~\eta_{\ChargeC\ParityP}\,\boldsymbol{a}^\dagger(-\vec p)\;.
\end{align} 
\end{subequations}
Here $\eta_{\ChargeC\ParityP}$ is a phase factor. As a consequence, scalar field operators transform as
\begin{equation}
 \left(\boldsymbol{\ChargeC}\,\boldsymbol{\ParityP}\right)^{-1}
 \,\boldsymbol{\phi}(x)\,
 \boldsymbol{\ChargeC}
 \,\boldsymbol{\ParityP}
 ~=~
 \eta_{\ChargeC\ParityP}\,\boldsymbol{\phi}^\dagger(\ParityP\,x)
\end{equation}
with $\ParityP\,x=(t,-\vec x)$. At the level of the Lagrangean, this
corresponds to a transformation 
\begin{equation}\label{eq:CanonicalCP4fields}
\phi(x)~\xmapsto{\boldsymbol{\ChargeC}\,\boldsymbol{\ParityP}}~
\eta_{\ChargeC\ParityP}\,\phi^*(\mathcal{P}x)
\end{equation}
for the fields, and we see that $\eta_{\ChargeC\ParityP}$ represents the
freedom of rephasing the fields. Analogous considerations for Dirac spinor
fields result in the transformation
\begin{equation}
\Psi(x)~\xmapsto{\boldsymbol{\ChargeC}\,\boldsymbol{\ParityP}}~
\eta_{\ChargeC\ParityP}\,\ChargeC^{T}\,\Psi^*(\mathcal{P}x)\;,
\end{equation}
where $\ChargeC$ is the charge conjugation matrix.

A Lagrangean, which is invariant under CPT, is
schematically given by 
\begin{equation}
\mathscr{L}~=~ c \, \mathcal{O}(x) + c^{\ast} \, \mathcal{O}^{\dagger}(x) \;,
\end{equation}
where $c$ is a coupling constant and $\mathcal{O}$ is an operator. Under
a physical CP transformation, 
\begin{equation}
\mathcal{O}(x) ~\xmapsto{\boldsymbol{\ChargeC}\,\boldsymbol{\ParityP}}~
\eta_\mathrm{CP}\,\mathcal{O}^{\dagger}(\mathcal{P}x)
\quad\text{and}\quad
c ~\xmapsto{\boldsymbol{\ChargeC}\,\boldsymbol{\ParityP}}~ c \;.
\end{equation}
{Demanding the Lagrangean to be invariant under the CP
transformation then restricts the phase of the coupling constant $c$.} 
In this case  the physical CP asymmetry of scattering amplitudes
\begin{equation}
 \varepsilon_{i\to f}~:=~
 \frac{\left|\Gamma(i\to f)\right|^2
 -\left|\Gamma(\bar{\imath}\to\bar{f})\right|^2 }{\left|\Gamma(i\to
  f)\right|^2+\left|\Gamma(\bar{\imath}\to\bar{f})\right|^2}
\end{equation}
will vanish to all orders in perturbation theory. Here $\bar{\imath}$ and
$\bar{f}$ denote the CP conjugate states of $i$ and $f$, and
are composed out of the corresponding anti--particles. As discussed above,
anti--particles are, per definition, related to the particles via \eqref{eq:fieldCP}.

\subsection{Generalizing CP transformations}
\label{sec:GeneralizingCP}

If the setting enjoys a discrete symmetry \DiscreteGroup, such that
$\boldsymbol{\phi}$ furnishes a non--trivial representation of \DiscreteGroup,
the phase factor $\eta_{\ChargeC\ParityP}$ in \eqref{eq:fieldCP} may (and,
as we shall see shortly, in general has to) be promoted to a unitary matrix
\UCP\ representing an automorphism transformation of \DiscreteGroup
\cite{Holthausen:2012dk}
\begin{subequations}\label{eq:GeneralizedCP}
\begin{align}
 \left(\boldsymbol{\ChargeC}
 \,\boldsymbol{\ParityP}\right)^{-1}\,\boldsymbol{a}(\vec p)\,\boldsymbol{\ChargeC}
 \,\boldsymbol{\ParityP}
 &~=~\UCP\,\boldsymbol{b}(-\vec p)
 \;, & 
 \left(\boldsymbol{\ChargeC}\,\boldsymbol{\ParityP}\right)^{-1}\,\boldsymbol{a}^\dagger(\vec p)\,\boldsymbol{\ChargeC}
 \,\boldsymbol{\ParityP}
 &~=~\boldsymbol{b}^\dagger(-\vec p)\,\UCP^\dagger
 \;,\\
 \left(\boldsymbol{\ChargeC}\,\boldsymbol{\ParityP}\right)^{-1}\,\boldsymbol{b}(\vec p)\,\boldsymbol{\ChargeC}
 \,\boldsymbol{\ParityP}
 &~=~\boldsymbol{a}(-\vec p)\,\UCP^\dagger
 \;, & 
\left(\boldsymbol{\ChargeC}\,\boldsymbol{\ParityP}\right)^{-1}\,\boldsymbol{b}^\dagger(\vec p)\,\boldsymbol{\ChargeC}
 \,\boldsymbol{\ParityP}
 &~=~\UCP\,\boldsymbol{a}^\dagger(-\vec p)\;,
\end{align} 
\end{subequations}
thus leading to a generalized CP transformation
\cite{Ecker:1981wv,Ecker:1983hz,Branco:1999fs}
\begin{equation}
 \phi(x)~\xmapsto{\CPgen}~
 \UCP\,\phi^*(\ParityP\,x)\;.
\end{equation}

Let us briefly explain, following Holthausen, Lindner and Schmidt
(HLS)~\cite{Holthausen:2012dk}, why it is necessary to generalize CP. Consider a
model based on the symmetry group \Tprime\ with two triplets $x$ and $y$ as well
as a field $\phi$ transforming as non--trivial one--dimensional representation
$\rep[_2]{1}$. Then the coupling (see \Appref{app:MaBasis} for our
conventions for \Tprime)
\begin{align}
 \left[\phi_{\rep[_2]{1}}\otimes
 \left(x_{\rep{3}}\otimes y_{\rep{3}}\right)_{\rep[_1]{1}}\right]_{\rep[_0]{1}} 
 & ~=~  \frac{1}{\sqrt{3}}\,\left[\phi\,\left(
 x_{1}\,y_{1}+\omega^2\,x_{2}\,y_{2}+\omega\,x_{3}\,y_{3}\right)\right]\;,
 \label{eq:toycoupling}
\end{align}
is \Tprime\ invariant.
A canonical CP transformation
\begin{equation}
 x~\xmapsto{\boldsymbol{\mathcal{C}\mathcal{P}}}~x^*
 \;,\quad
 y~\xmapsto{\boldsymbol{\mathcal{C}\mathcal{P}}}~y^*\;,
 \quad\text{and}\quad
 \phi~\xmapsto{\boldsymbol{\mathcal{C}\mathcal{P}}}~\phi^*\;,
\label{eq:canonicalCP}
\end{equation}
would map the $\rep[_2]{1}$ representation $\phi$ to a $\rep[_1]{1}$ such that
the contraction  \eqref{eq:toycoupling} gets mapped to a term which
is not \Tprime\ invariant. This can be repaired by imposing a generalized CP
transformation $\CPgen$, which we
discuss in more detail later in \Secref{sec:twoA}, and under which
\begin{equation}
 \left(\begin{array}{c}x_1\\ x_2\\ x_3\end{array}\right)
 ~\xmapsto{\CPgen}~
 \left(\begin{array}{c}x_1^*\\ x_3^*\\ x_2^*\end{array}\right)
 \;,\quad
 \left(\begin{array}{c}y_1\\ y_2\\ y_3\end{array}\right)
 ~\xmapsto{\CPgen}~
 \left(\begin{array}{c}y_1^*\\ y_3^*\\ y_2^*\end{array}\right)\;,
 \quad\text{and}\quad
 \phi~\xmapsto{\CPgen}~\phi^*
 \;.\label{eq:toyCPtilde}
\end{equation}
Under this transformation, the contraction \eqref{eq:toycoupling} gets mapped to
its hermitean conjugate. The Lagrangean then respects the generalized CP
symmetry if the coupling coefficient is real. The crucial property of the
transformation \eqref{eq:toyCPtilde} is that it is not composed of a canonical
CP and a \Tprime\ symmetry transformation. Rather, it involves an outer
automorphism of this group \cite{Holthausen:2012dk}.

The heart of the above problem seems to be related to the complexity of the
Clebsch--Gordan (CG) coefficients appearing in \Eqref{eq:toycoupling}. One may
then speculate that one might have to switch to a basis in which all CG's are real,
and impose the canonical CP transformation there. The purpose of our discussion
is to show that the true picture is somewhat more subtle. First of all, we
will see that there are groups which do not admit real CG's but nevertheless
allow for a consistent CP transformation, which, if it is a symmetry of the
Lagrangean, ensures physical CP conservation. Second, we shall show that there
are symmetry groups that do not allow for a transformation which ensures
physical CP conservation. 
The CG's in such groups are always complex, and models based on such symmetries
will, at least generically, violate CP. In other words, for such groups CP
violation originates from group theory~\cite{Chen:2009gf}, thus providing us
with very interesting explanation for why CP is violated in Nature relating CP
violation to flavor.

\subsection{What are the proper constraints on generalized CP transformations?}
\label{sec:ProperCP}

Let us now discuss the general properties of generalized CP transformations. As
discussed in \HLS\ (see also \cite{Feruglio:2012cw}) and above, generalized CP
transformations are given by  automorphisms of the group \DiscreteGroup, since
otherwise the transformation would map \DiscreteGroup--invariant terms in the
Lagrangean to non--invariant terms.

However, the only way to generalize CP in a model--independent way is to demand
that the operators $\boldsymbol{a}$ and $\boldsymbol{b}$ in
\eqref{eq:GeneralizedCP} get interchanged. Imposing a ``generalized CP
transformation''  that does not have this property will, in general, not warrant
physical CP conservation.  This is because it does not map field operators to
their \textit{own} hermitean  conjugates. In fact, as we shall discuss in an
explicit example  (see~\Secref{sec:CPlike}), such a ``generalized CP symmetry''
does not lead to a vanishing decay asymmetry. That is, in models with very
specific field content one may re--define CP  such that it contains a
non--trivial interchange of fields in representations which  are not related by
complex conjugation. The violation of the thus ``generalized CP'' is
then, however, no longer a prerequisite for, say,
baryogenesis.  We therefore prefer to refer to such transformations as
``CP--like'' transformations.   As we are interested in the origin of physical
CP violation, we will from now on impose  that the operators $\boldsymbol{a}$
and $\boldsymbol{b}$ in \eqref{eq:GeneralizedCP} get interchanged.  This implies
that a true (generalized) CP transformation has to map all complex (irreducible)
representations of \DiscreteGroup\ to their conjugates.

Let us now discuss CP transformations that generalize the canonical CP
transformation \eqref{eq:CanonicalCP4fields} and act on scalar fields as
\begin{equation}
\label{eq:generalized_CP}
 \Phi(x)  ~\xmapsto{\CPgen}~U_\mathrm{CP}\,\Phi^*(\mathcal{P}\,x)\;,
\end{equation}
where $U_\mathrm{CP}$ is a unitary matrix and $\Phi$ contains, in principle, all
fields of the model. Here and in what follows, we will only discuss the
transformation of scalar fields; the extension to higher--spin fields is
straightforward.  \HLS\ showed that this generalized CP transformation is only
consistent with the flavor symmetry group \DiscreteGroup\ if \UCP\ is
non--trivially related to an automorphism $u~:~\DiscreteGroup \to
\DiscreteGroup$. In fact, \UCP\ has to be a solution to the consistency equation
(cf.\ equation~(2.8) in \HLS\ and see also \cite{Feruglio:2012cw})
\begin{equation}\label{eq:consistencyHLS}
  \rho\bigl(u(g)\bigr)
  ~=~
  \UCP\,\rho(g)^*\,\UCP^\dagger  \quad\forall~g\in\DiscreteGroup\;,
\end{equation}
where $\rho(g)$ is the (in general reducible) matrix representation in which
$\Phi$ transforms under \DiscreteGroup. However, if \Eqref{eq:generalized_CP} is
to be a physical CP transformation, $u$ has to have, in generic settings, some
further properties:

\paragraph{$\boldsymbol{u}$ has to be class--inverting.} As discussed above,
 we demand that $u$ maps every irreducible representation \rep[_i]{r} to its
\emph{own} conjugate. Therefore, the matrix realizations
 $\rhoR{_i}$ fulfill
 \begin{equation}\label{eq:consistency1}
  \rhoR{_i}\!\bigl(u(g)\bigr)~=~\UU[_i]{\rep{r}}\,\rhoR{_i}\!(g)^*\,\UU[_i]{\rep{r}}^\dagger
  \quad\forall~g\in\DiscreteGroup~\text{and}~\forall~i\;,
 \end{equation}
 with some unitary matrices $\UU[_i]{\rep{r}}$. This implies, of course, that (pseudo--)real representations
 get mapped to themselves. The matrix \UCP\ of \Eqref{eq:generalized_CP} is
 given by the direct sum of the $\UU[_i]{\rep{r}}$ corresponding to the particle
 content of the model at hand, or, more explicitly, \UCP\ is composed of  blocks
 consisting of the $\UU[_i]{\rep{r}}$,
\begin{align}
 \Phi~=~
  \left(\begin{array}{c}\uparrow\\ \phi_{\rep[_{i_1}]{r}}\\ \downarrow \\ \hline 
 \uparrow\\ \phi_{\rep[_{i_2}]{r}}\\ \downarrow \\ \hline \vdots\end{array}\right)
 ~\xmapsto{\CPgen}~&
 \left(\begin{array}{ccc|ccc|c}
 \nwarrow & & \nearrow & & & & \\
 &\!\!\! \UU[_{i_1}]{r} \!\!\!&  & & & & \\
 \swarrow & & \searrow & & & & \\
 \hline
 & & & \nwarrow  & & \nearrow &  \\
 & & & & \!\!\!\UU[_{i_2}]{r}\!\!\!  & & \\
 & & & \swarrow  & & \searrow &  \\
 \hline 
 & & & & & & \ddots
 \end{array}\right)
 \,
 \left(\begin{array}{c}\uparrow\\ \phi_{\rep[_{i_1}]{r}}^*\\ \downarrow \\ \hline 
 \uparrow\\ \phi_{\rep[_{i_2}]{r}}^*\\ \downarrow \\ \hline \vdots\end{array}\right)
 \notag\\
 ~=~&\UCP\,\Phi^*
 \;,
\end{align} 
where $\phi_{\rep[_{i_a}]{r}}$ transforms in representation \rep[_{i_a}]{r}.
Clearly, the precise form of $\UCP$ depends on the model, yet the
$\UU[_i]{\rep{r}}$ depend on the symmetry \DiscreteGroup\ only. This allows us
to define a CP transformation for a discrete symmetry \DiscreteGroup\ rather
than for a given model with a particular representation content. In this
point, our discussion differs from the one in \HLS, where \UCP\ is allowed not
to be block--diagonal.

Further, taking the trace reveals that the group characters $\chiR{_i}$ fulfill
\begin{align}
  \chiR{_i}\!\left(u(g)\right)
  &~=~
  \tr\left[\rhoR{_i}\!\left(u(g)\right)\right]
  ~=~
  \tr\left[\UU[_{i}]{r}\,\rhoR{_i}\!(g)^*\,\UU[_{i}]{r}^\dagger\right]
  \notag\\
  &~=~
 \tr\left[\rhoR{_i}\!(g)\right]^*
  ~=~
  \chiR{_i}\!(g)^*
  ~=~
  \chiR{_i}\!(g^{-1})\quad \forall~i\;,
\end{align}
i.e.\ $u$ is class--inverting.\footnote{A class--inverting automorphism
$u$ sends each group element $g$ to an element $u(g)$ which lies in the
same conjugacy class as $g^{-1}$, i.e.\ $u(g)=h\,g^{-1}\,h^{-1}$ for some $h
\in \DiscreteGroup$.}

\paragraph{Comments on the order of $\boldsymbol{u}$.} 
 Under the square of the generalized CP transformation, $\Phi$ transforms as
 \begin{equation} 
  \Phi~\xmapsto{\CPgen^2}~\UCP\,\left(
  \UCP\,\Phi^*(\mathcal{P}^2\,x)\right)^*
  ~=~\UCP\,\UCP^*\,\Phi(x)~=:~V\,\Phi(x)
 \end{equation}
 with some unitary matrix $V$ which can be related to the automorphism $v=u^2$.
 Since \UCP\ is a matrix direct sum, one can again discuss the different
 irreducible representations of
 \DiscreteGroup\ separately,
 \begin{equation} 
  \phi_{\rep[_i]{r}}~\xmapsto{\CPgen^2}~\UU[_i]{\rep{r}}\,\left(
  \UU[_i]{\rep{r}}\,\phi_{\rep[_i]{r}}^*(\mathcal{P}^2\,x)\right)^*
  ~=~\UU[_i]{\rep{r}}\,\UU[_i]{\rep{r}}^*\,\phi_{\rep[_i]{r}}(x)~=:~V_{\rep[_i]{r}}\,\phi_{\rep[_i]{r}}(x)\qquad \forall~i\;.
 \end{equation}
Imposing the CP transformation of \eqref{eq:generalized_CP} as a symmetry
immediately implies that $\Phi \to V\,\Phi$ is also a symmetry transformation.
Note that, as the square of a class--inverting automorphism, $v=u^2$
is class--preserving. One can distinguish now three cases:
\begin{itemize}
 \item[(i)]~$u$ is involutory, i.e.\ $u^2=v=\text{identity (id)}$,
 \item[(ii)]~$v=u^2$ is an inner automorphism, and
 \item[(iii)]~$v=u^2$ is an outer automorphism\footnote{Note that there are
class--preserving automorphisms that are not inner automorphisms.}
\end{itemize}
which we will discuss in the following.

Let us start with case~(i), where the order of the automorphism $u$ is at most
two. We will now show that if and only if this is the case, the matrices
$V_{\rep[_i]{r}}$ are $\pm\mathbbm{1}$. 

First we start with the consistency condition \Eqref{eq:consistency1} for a
class--inverting $u$. By replacing $g$ by $u(g)$ (and $u(g)$ by $u^2(g)$) in
\Eqref{eq:consistency1} and bringing the $\UU[_i]{\rep{r}}$'s to the other
side, we obtain 
\begin{equation}
  \rhoR{_i}\!(u(g)) ~=~ \UU[_i]{\rep{r}}^T\, \rhoR{_i}\!(u^2(g))^* \,\UU[_i]{\rep{r}}^* 
  ~=~ \UU[_i]{\rep{r}}^T\, \rhoR{_i}\!(g)^* \,\UU[_i]{\rep{r}}^* \quad\forall~g\in\DiscreteGroup~\text{and}~\forall~i\;.
\end{equation}
This shows that with $\UU[_i]{\rep{r}}$ also the transpose $\UU[_i]{\rep{r}}^T$
fulfills \Eqref{eq:consistency1}. Since $\rhoR{_i}$ is an irreducible
representation, Schur's Lemma implies that
 \begin{equation}\label{eq:U(anti)symmetric}
   \UU[_i]{\rep{r}}^T ~=~ \mathrm{e}^{\I\, \alpha} \, \UU[_i]{\rep{r}} \qquad \forall~i\;,
 \end{equation}
which is only possible if each $\UU[_i]{\rep{r}}$ is either symmetric or
anti--symmetric, i.e.\  $\alpha$ is either $0$ or $\pi$. Thus,
$V_{\rep[_i]{r}}=\UU[_i]{\rep{r}}\,\UU[_i]{\rep{r}}^*=\pm\mathbbm{1}$. Hence,
$V$ consists of blocks identical to $\pm\mathbbm{1}$.  
 
Now assume that all $V_{\rep[_i]{r}}$ are $\pm\mathbbm{1}$. Then, by
inserting \Eqref{eq:consistency1} into itself,
\begin{equation}
 \rhoR{_i}\!(u^2(g)) ~=~ \left(\UU[_i]{\rep{r}}\,\UU[_i]{\rep{r}}^*\right) \, 
 \rhoR{_i}\!(g) \, \left(\UU[_i]{\rep{r}}\,\UU[_i]{\rep{r}}^*\right)^\dagger 
 ~=~ \rhoR{_i}\!(g) \qquad \forall\, g\in\DiscreteGroup~\text{and}~\forall~i\;.
\end{equation}
Since this equation is true for all irreducible representations, it follows
that $u^2(g)=g$ for all $g$ in \DiscreteGroup\ and the order of $u$ is
thus either one or two (i.e.\ $u$ is involuntary). This completes
the proof that $V$ is different from a diagonal matrix with only
$\pm\mathbbm{1}$ on the diagonal if and only if $u$ is of order $n>2$.
  
We therefore conclude that, if an involutory $u$ is imposed as a symmetry, 
\DiscreteGroup\ may be amended by an additional \Z2 symmetry. This is
possible if and only if $V_{\rep[_i]{r}}\neq +\mathbbm{1}$ for some
$\rep[_i]{r}$. We will discuss this case in more detail in
\Secref{sec:TypeIICPtrafo}. In what follows, we refer to such an enlargement as ``trivial''
extension of \DiscreteGroup\ to $\DiscreteGroup\times\Z2$. Note that the assignment of \Z2 charges to the fields of a model
is not arbitrary but is given by the signs of the $V_{\rep[_i]{r}}$ for their
respective representations under \DiscreteGroup. We will also discuss this \Z2
factor in an example in~\Secref{sec:twoB}.

The second logical possibility, case (ii), is that $v$ is an inner
automorphism.\footnote{The property that $u$ should square to the identity or an
inner automorphism has also been stressed in \cite{Nishi:2013jqa}. However, the
discussion there misses the point that this does not imply
$\UU[_i]{\rep{r}}\UU[_i]{\rep{r}}^*=\rhoR{_i}(g)$ for some $g$ in
\DiscreteGroup\ but that the group still might be extended by a \Z2 factor.} In
this case, the order of $u$ is larger than two but one can still show that the
flavor group only gets enlarged by some Abelian factor. However, CP
transformations connected to automorphisms that square to an inner automorphism
do not seem to yield any CP transformations which are physically different from
those that are connected to involutory automorphisms. The reason is that if two
automorphisms $u$ and $u'$ are related by an inner automorphism,
\begin{equation}
 u(g) ~=~ b\,u'(g)\,b^{-1} \quad \forall ~g \in \DiscreteGroup
 ~\text{and some}~b\in \DiscreteGroup\;,
\end{equation}
the resulting CP transformations only differ by a transformation with the group
element $\rho(b)$.  Since the latter transformation certainly is a symmetry of
the Lagrangean, the two CP transformations are indistinguishable. In fact, it
turns out that we were not able to find an example where there is a
class--inverting automorphism of higher order that is not related to an
involutory class--inverting automorphism in the prescribed way. We were able to
prove that such an automorphism cannot exist for some cases, see
\Appref{app:ClassInverting}, and have explicitly checked this for all
non--Abelian groups of order less than 150 (with the exception of some groups of
order 128) with the group theory program GAP~\cite{GAP4}.

The last logical possibility, case (iii), is that $u^2=v$ is an outer
automorphism. Then the additional generator $h$ with
$\rhoR{_i}(h)=V_{\rep[_i]{r}}$ does not commute with all group elements of
\DiscreteGroup, and, hence, genuinely enlarges the original flavor symmetry
group non--trivially to the larger semi--direct product group 
$H=\DiscreteGroup\SemiDirect_{v} \Z{h}$, where $\Z{h}$ is the cyclic group
generated by $h$.  As a consequence, terms which are allowed by \DiscreteGroup\
but prohibited by $H$ are absent from a Lagrangean if CP conservation is
imposed. The representation content, however, still coincides with the one of
\DiscreteGroup.  Although the structure of the extended group $H$ is more
complicated than the direct product in case (i), the physical implications are
similar to the case of the trivial \Z2 extension.

Even though we have no general argument for their absence, we were not able to
find an example in which case (iii) is realized. In more detail, a GAP
scan for class--preserving outer automorphisms that are
the square of a class--inverting inner automorphism did not yield any result for
groups up to order 150 (some groups of order 128 were not checked). Case
(iii), hence, seems to be very rare. 

In summary, we find that $u$ should be a class--inverting  automorphism of
\DiscreteGroup\ in order for the related CP transformation to be physical. 
Moreover, for practical purposes, one can usually restrict the discussion to
involutory automorphisms.

\subsection{The Bickerstaff--Damhus automorphism (BDA)}
\label{sec:BDA}

As shown by Bickerstaff and Damhus \cite{Bickerstaff:1985jc}, the existence of a
basis in which all CG coefficients are real can be related to the existence
of an automorphism $u$ which fulfills
\begin{equation}\label{eq:BDAequation}
 \rhoR{_i}\!\left(u(g)\right)~=~\UU[_i]{\rep{r}}\,\rhoR{_i}\!(g)^*\,\UU[_i]{\rep{r}}^\dagger\;,\quad 
 \UU[_i]{\rep{r}}~\text{unitary and symmetric,}\quad\forall~g\in\DiscreteGroup
 ~\text{and}~\forall~i
\end{equation}
for some $\UU[_i]{\rep{r}}$ with the given properties.  From our discussion in
\Secref{sec:ProperCP} we know that such a $u$ is involutory and
class--inverting. In what follows, we will refer to an automorphism $u$ which
satisfies \Eqref{eq:BDAequation} as Bickerstaff--Damhus automorphism (BDA). In
short, a BDA is a class--inverting involutory automorphism that fulfills the
consistency condition \eqref{eq:consistency1} with some symmetric unitary
matrices $\UU[_i]{\rep{r}}$.

The important property of the BDA is that its existence is equivalent to the
existence of a basis of \DiscreteGroup\ in which all CG's are real,
\begin{equation}
 \exists~\text{BDA}~u~\text{fulfilling}~(\ref{eq:BDAequation})
 \quad
 \Longleftrightarrow
 \quad
 \left\{\begin{array}{c}\text{existence of a}\\ \text{basis in which}\\
 \text{all CG coefficients}\\ \text{are real}\end{array}\right\}\;.
\end{equation}
The basis in which the CG's can be chosen real is exactly the basis for which
all $\UU[_i]{\rep{r}}$ in \Eqref{eq:BDAequation} are unit matrices, i.e.\ for which
\begin{equation}\label{eq:BDAequation2}
 \rhoR{_i}\!\left(u(g)\right)~=~\rhoR{_i}\!(g)^*\quad\forall~g\in\DiscreteGroup ~\text{and}~\forall~i\;.
\end{equation}
More precisely, this defines a whole set of bases which are related by real
orthogonal basis transformations.

An automorphism $u$ that fulfills this equation in a certain basis is unique.
However, there can be several different BDAs which fulfill
\Eqref{eq:BDAequation2} for different bases.  The different BDAs also do not
have to be related by inner automorphisms, see for example the group
$\text{SG}(32,43)$ of the SmallGroups library which is part of GAP.  Note that,
as shown in \Appref{app:proofOddOrder}, odd order non--Abelian groups do not
admit a BDA and, hence, do not have a basis with completely real Clebsch--Gordan
coefficients.

How can one tell whether or not a given automorphism $u$ is a BDA? In what follows, we will discuss a tool which allows us to answer this
question.

\subsection{The twisted Frobenius--Schur indicator}

The Frobenius--Schur indicator (see e.g.~\cite[p.~48]{Ramond:2010zz}) is a
well--known tool to distinguish real, pseudo--real, and complex representations
of a finite group. It is defined by
\begin{equation}
  \mathrm{FS}(\rep[_i]{r}) ~:=~ 
  \frac{1}{|\DiscreteGroup|}\,\sum_{g \in \DiscreteGroup} \, 
        \chiR{_i}\!(g^2) 
  ~=~ \frac{1}{|\DiscreteGroup|}\,\sum_{g \in \DiscreteGroup} \,
        \tr{\left[\rhoR{_i}\!(g)^2\right]}\; ,
\end{equation}
with $|\DiscreteGroup|$ being the order of the group $\DiscreteGroup$.
The result is
\begin{equation}
  \mathrm{FS}(\rep[_i]{r}) ~=~ 
  \left\{\begin{array}{ll}
    +1,& \text{~if \rep[_i]{r} is a real representation,}\\
    0, & \text{~if \rep[_i]{r} is a complex representation,}\\
    -1,& \text{~if \rep[_i]{r} is a pseudo--real representation.}
  \end{array}\right.
\end{equation}

\begin{figure}[h]
\centerline{\includegraphics{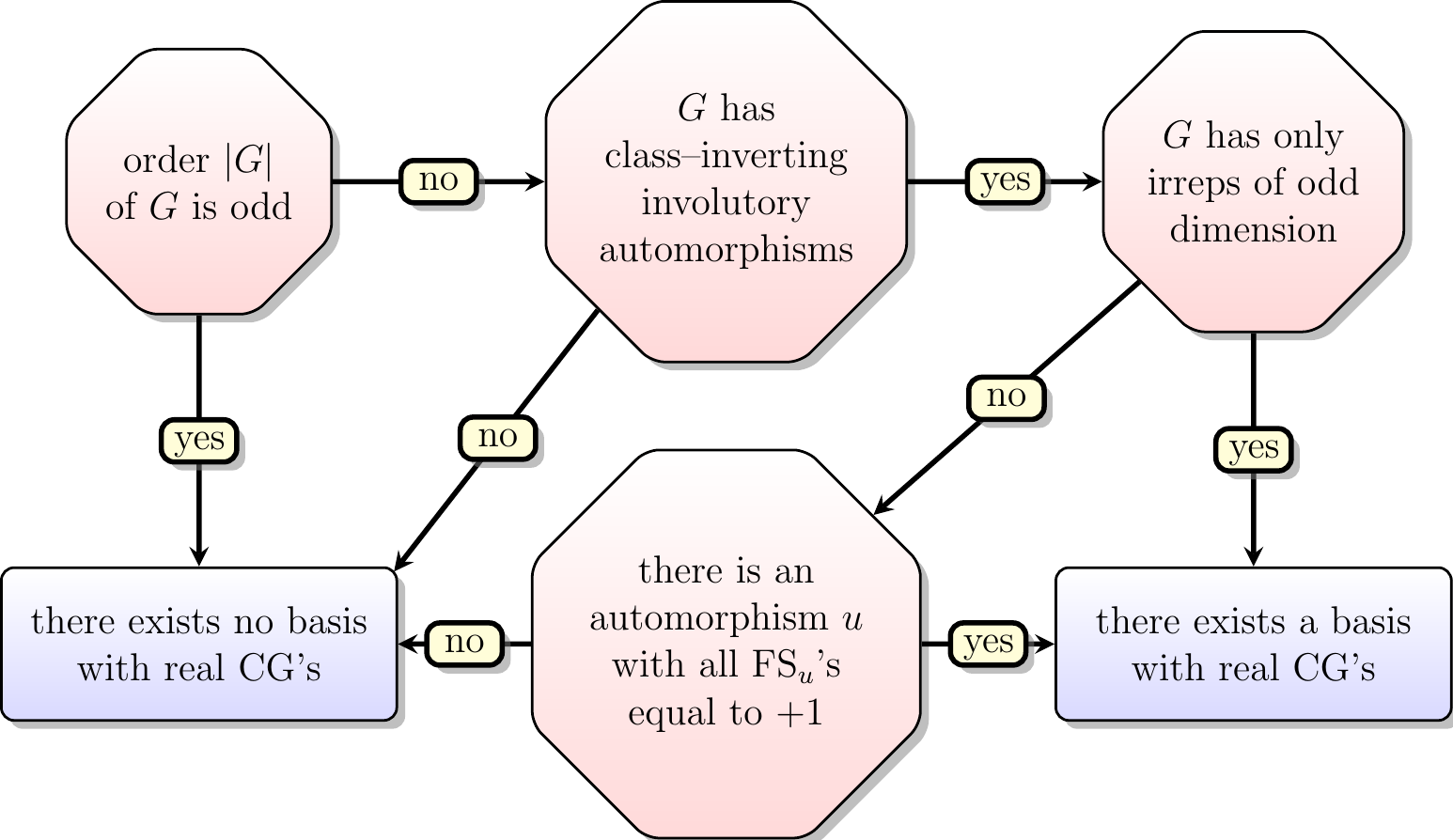}}
  \caption{This flowchart displays a possible sequence of steps one could follow
to determine whether a finite non--Abelian group \DiscreteGroup\ admits a basis with real
Clebsch--Gordan coefficients.}
  \label{fig:RealCGs} 
\end{figure}

In complete analogy to the Frobenius--Schur indicator, one can define the
twisted Frobenius--Schur indicator (\FSI) \cite{Bickerstaff:1985jc,Kawanaka1990} that depends on an automorphism $u$ and that
determines whether $u$ is a Bickerstaff--Damhus
automorphism. In fact, for an automorphism $u$ we will show that
\begin{equation}\label{eq:FSI}
  \FSI(\rep[_i]{r}) ~=~ 
  \left\{\begin{array}{ll}
    +1 \quad \forall~i, &~\text{if}~u\text{~is a Bickerstaff--Damhus automorphism,}\\
    +1~\text{or}~-1 \quad \forall~i,& ~\text{if}~u\text{~is class--inverting and involutory,}\\
    \text{different from $\pm1$},& ~\text{if}~
        u\text{~is not class--inverting and/or not involutory.}
  \end{array}\right.
\end{equation}
Our recipe for determining whether or not a finite non--Abelian group
\DiscreteGroup\ admits a basis with real Clebsch--Gordan coefficients,
using the twisted Frobenius--Schur indicator, is outlined in
\Figref{fig:RealCGs}.

The twisted Frobenius--Schur indicator for an irreducible representation
\rep[_i]{r} and an automorphism $u$ is defined as
\begin{align}
  \FSI(\rep[_i]{r}) ~:=~ 
  \frac{1}{|\DiscreteGroup|}\,\sum_{g \in \DiscreteGroup} \, 
        \chiR{_i}\!(g \, u(g)) 
  & ~=~ \frac{1}{|\DiscreteGroup|}\,\sum_{g \in \DiscreteGroup} \,
        \tr{\left[\rhoR{_i}\!(g)\,\rhoR{_i}\!(u(g))\right]}\notag\\
  & ~=~ \frac{1}{|\DiscreteGroup|}\,\sum_{g \in \DiscreteGroup} \, 
  \left[\rhoR{_i}\!(g)\right]_{\alpha\beta} \, 
  \left[\rhoR{_i}\!(u(g))\right]_{\beta\alpha}\;,
\end{align}
where we sum over the matrix indices $\alpha$ and $\beta$.
From the definition it is immediately clear that for $u\equiv\mathrm{id}$ one
recovers the original Frobenius--Schur indicator.

The proof of the statements in \Eqref{eq:FSI} is based on the well--known Schur
orthogonality relation for the irreducible representations \rep[_i]{r} of the
group \DiscreteGroup\ (see e.g.~\cite[p.~37]{Ramond:2010zz}),
\begin{equation}\label{eq:Schurorthogonality}
  \sum_{g \in \DiscreteGroup} \, \left[\rhoR{_i}\!(g)^*\right]_{\alpha\beta} \,
  \left[\rhoR{_j}(g)\right]_{\gamma\delta} ~=~ \frac{|\DiscreteGroup|}{\dim
  \rep[_i]{r}}\,\delta_{ij}\,\delta_{\alpha\gamma}\,\delta_{\beta\delta}\;.
\end{equation}

The irreducible representations realized by $\rhoR{_i}\!(g)$ and
$[\rhoR{_i}\!(u(g))]^*$ are equivalent for all $i$ if and only if $u$
is class--inverting. Hence, if $u$ is not class--inverting, according to
\Eqref{eq:Schurorthogonality}, the twisted Frobenius--Schur indicator vanishes
for at least one irreducible representation.

Let now $u$ be class--inverting. Then there is a unitary matrix $\UU[_i]{\rep{r}}$ for each
irreducible representation \rep[_i]{r} such that
\begin{equation}\label{eq:conjugateEquivalence}
  \rhoR{_i}\!(u(g)) ~=~ \UU[_i]{\rep{r}} \, \rhoR{_i}\!(g)^* \, \UU[_i]{\rep{r}}^\dagger\;, \qquad \forall ~ i \;.
\end{equation}
Inserting this into the twisted Frobenius--Schur indicator and simplifying the
expression, one arrives at
\begin{align}
  \FSI(\rep[_i]{r}) & 
  ~=~ \frac{1}{|\DiscreteGroup|}\,\sum_{g \in \DiscreteGroup} \, 
  \left[\rhoR{_i}\!(g)\right]_{\alpha\beta} \, \left[\UU[_i]{\rep{r}}\right]_{\beta\gamma} 
  \, \left[\rhoR{_i}\!(g)^*\right]_{\gamma\delta} \, \left[\UU[_i]{\rep{r}}^\dagger\right]_{\delta\alpha}\notag\\
  & \stackrel{(\ref{eq:Schurorthogonality})}{~=~} 
  \frac{1}{|\DiscreteGroup|}\, \left[\UU[_i]{\rep{r}}\right]_{\beta\gamma} \, \left[\UU[_i]{\rep{r}}^\dagger\right]_{\delta\alpha} \, \frac{|\DiscreteGroup|}{\dim \rep[_i]{r}}\,\delta_{\alpha\gamma}\,\delta_{\beta\delta}\notag\\
  & ~=~ \frac{1}{\dim \rep[_i]{r}}\,\tr{\left(\UU[_i]{\rep{r}} \, \UU[_i]{\rep{r}}^*\right)} 
  ~=~ \frac{1}{\dim \rep[_i]{r}}\,\tr{\left(V_{\rep[_i]{r}}\right)}\;.
\end{align}
As shown in \Secref{sec:ProperCP}, $V_{\rep[_i]{r}}$ is $\pm \mathbbm{1}$ if and
only if $u$ is an involution, where plus signals a symmetric and minus an
anti--symmetric matrix $\UU[_i]{\rep{r}}$. Hence, if and only if $u$ is a
class--inverting involution, the twisted Frobenius--Schur indicator is $\pm1$
for all irreps \rep[_i]{r}. Furthermore, $u$ is a Bickerstaff--Damhus
automorphism if and only if \Eqref{eq:BDAequation} holds with symmetric matrices
$\UU[_i]{\rep{r}}$. Thus, $u$ is a BDA if and only if the twisted
Frobenius--Schur indicators of all irreducible representations of
\DiscreteGroup\ are $+1$. This completes the proof of \Eqref{eq:FSI}.

It is important to note that the twisted Frobenius--Schur indicator can vanish
for higher--order automorphisms even though they are class--inverting. For such
automorphisms, one can define an extended version of the indicator, which again
has the property to be $\pm1$ for all irreps in the class--inverting case and
$0$ for some irrep otherwise. Let $n=\ord{(u)}/2$ for even--order and
$n=\ord{(u)}$ for odd--order automorphisms. Then the $n^\mathrm{th}$
extended\footnote{The $1^\mathrm{st}$ extended twisted Frobenius--Schur
indicator $\FSI^{(1)}$ is identical to the regular twisted \FSI.} twisted
Frobenius--Schur indicator
\begin{equation}\label{eq:FSIn}
  \FSI^{(n)}(\rep[_i]{r}) ~:=~ \frac{(\dim{\rep[_i]{r}})^{n-1}}{|\DiscreteGroup|^n}\,
  \sum_{g_1,\dots,g_n \in \DiscreteGroup} \, \chiR{_i}
  \bigl(g_1 \, u(g_1)\cdots g_n \, u(g_n)\bigr)
\end{equation}
is $\pm1$ for all irreducible representations if $u$ is class--inverting and $0$
for at least one irrep if not. A proof of this statement is given in
\Appref{app:extendedFS}.

\subsection{Three types of groups}
\label{sec:threeTypes}
  
\begin{figure}[!h!]
\centerline{\includegraphics{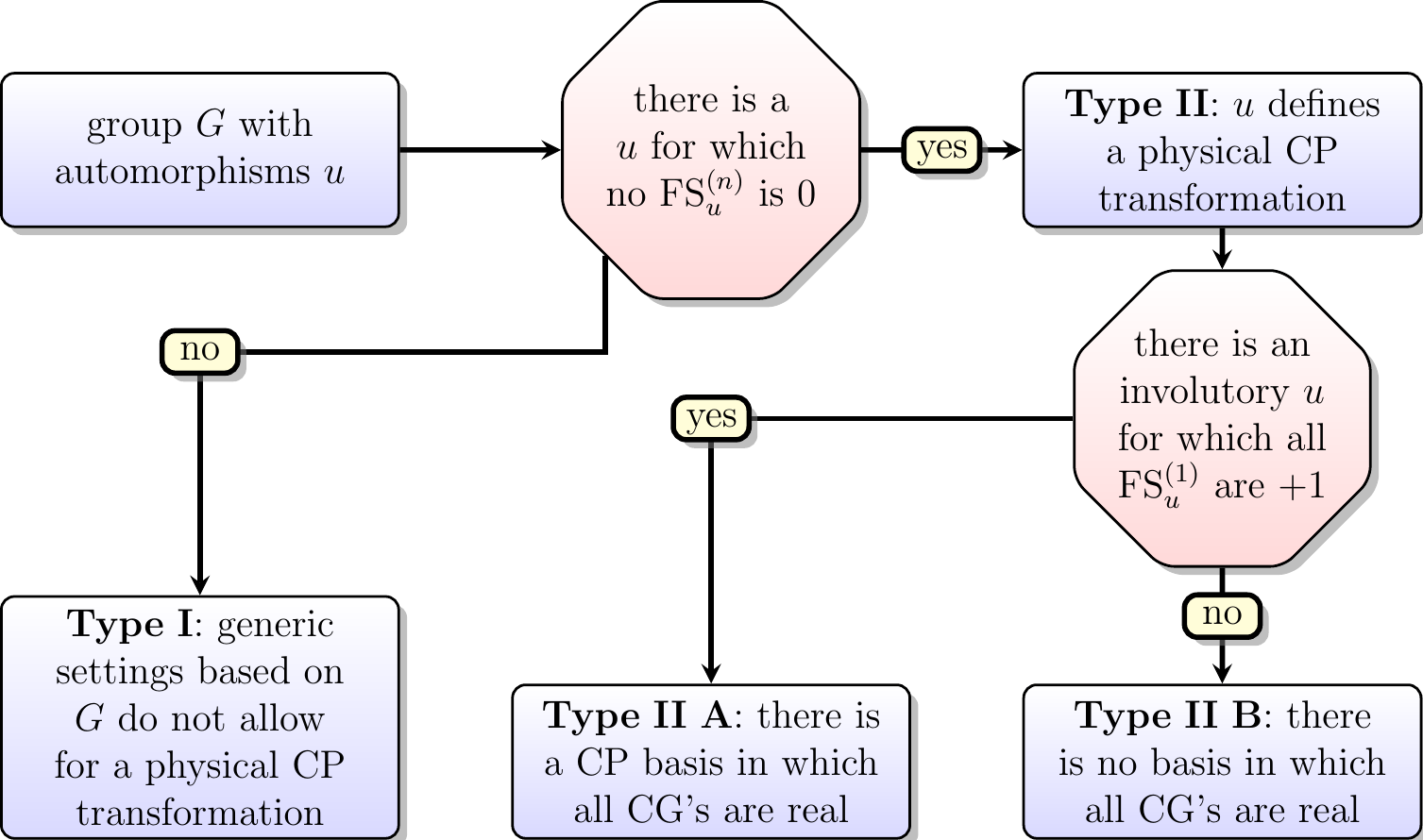}}
  \caption{This flowchart displays how the regular and extended twisted
   Frobenius--Schur indicators \FSI\ and $\FSI^{(n)}$ allow us to distinguish
   between the three types of groups. The integer $n$ is $n=\ord(u)/2$ for even
and $n=\ord(u)$ for odd $\ord(u)$.}
  \label{fig:3types} 
\end{figure}

The twisted Frobenius--Schur indicator can be used to categorize finite groups
into three classes. In order to do so, one has to compute the indicator for all
involutory automorphisms $u_\alpha$ of the specific finite group
\DiscreteGroup.\footnote{More precisely, one would have to calculate the
$\FSI^{(n)}$'s for all automorphisms. The
difference, however, is only relevant for the categorization if all
class--inverting automorphisms of \DiscreteGroup\ square to a non--trivial
outer automorphism. Groups in which this is the case would be classified as
type~II~B. However, an extensive scan (cf.\ \Secref{sec:ProperCP}) for such
groups did not yield any result. On the other hand, we were also not able to
prove that such groups cannot exist.} A code for the group theory software
GAP~\cite{GAP4} that performs this task is shown in \Appref{app:GAP}. There are
then three cases:
\begin{description}
 \item[Case~I:] for all involutory automorphisms $u_\alpha$ of \DiscreteGroup\
 there exists at least one representation $\rep[_i]{r}$ for which
 $\text{FS}_{u_\alpha}(\rep[_i]{r})=0$. In this case, the discrete symmetry
 \DiscreteGroup\ does not allow us to define a proper CP transformation in a
 generic setting.
 \item[Case~II:] for (at least) one involutory automorphism $u$ of
 \DiscreteGroup, the \FSI's for all representations are non--zero.  Then
 there are two sub--cases:
 \begin{description}
  \item[Case~II~A:] all \FSI's are $+1$ for one of those $u$'s.  Then this $u$
  is a BDA and there exists a basis with real Clebsch--Gordan coefficients. $u$
  can be used to define a proper CP transformation in any basis.\footnote{Note
  that these groups can have additional class--inverting involutory
  automorphisms that are not BDAs.}
  \item[Case~II~B:] some of the \FSI's are $-1$ for all such $u$'s.  Then
  there is no BDA, and, hence, one cannot find a basis in which all CG's are
  real. Yet any of these $u$'s can be used to define a proper CP
  transformation.  
 \end{description}
\end{description}
Depending on which case applies to \DiscreteGroup, we will from now on refer to
\DiscreteGroup\ as being of type~I, type~II~A and type~II~B, respectively (see
\Figref{fig:3types}).

In \Tabref{tab:groupTypes}, we list for each of the types several examples.
\begin{table}[!h!]
\begin{center}
\subtable[Examples for type I groups. Generally, all odd order non--Abelian
groups are of this type with the caveat of groups that have a class--inverting
automorphism that squares to a non--trivial outer one.
\label{tab:typeI}]{\begin{tabular}{|r|c|c|c|c|}
  \hline
  group & $\Z5 \SemiDirect \Z4$ & $T_7$  & $\Delta(27)$ & $\Z9 \SemiDirect \Z3$ \\
  \hline
  SG & (20,3) & (21,1) & (27,3) & (27,4) \\
  \hline
  \end{tabular}}\\
\subtable[Examples for type II~A groups. The dihedral and all
Abelian groups are also of this type.\label{tab:typeIIA}]{\begin{tabular}{|r|c|c|c|c|c|c|c|c|}
  \hline
  group 
  & $S_3$ & $Q_8$ & $A_4$ & $\Z3\rtimes\Z8$ & \Tprime &  $S_4$ & $A_5$\\
  \hline
  SG & (6,1) & (8,4) & (12,3) & (24,1) & (24,3) & (24,12) & (60,5)\\
  \hline
  \end{tabular}}\\
\subtable[Examples for type II~B groups.\label{tab:typeIIB}]{\begin{tabular}{|r|c|c|}
  \hline
  group 
  & $\Sigma(72)$ & $\left((\Z3 \times \Z3) \SemiDirect \Z4\right) \SemiDirect \Z4$\\
  \hline
  SG & (72,41) & (144,120) \\
  \hline
  \end{tabular}}
\end{center}  
\caption{Examples for the three types of groups: (a)~I, (b)~II~A and (c)~II~B
with their common names and SmallGroups library ID of GAP~\cite{GAP4}. }
\label{tab:groupTypes}
\end{table}

Let us also comment that, when building a concrete model, one may still be
able to define a proper CP transformation even in the type~I case by not introducing
any representation which has a zero \FSI. That is, groups of type I generically
violate CP, but physical CP violation is not guaranteed in non--generic models.
We will explain this statement in more detail in \Secref{sec:Delta27}.

\subsection{Physical CP transformations for type II groups}
\label{sec:TypeIICPtrafo}

Let us now discuss in more detail the proper physical CP transformations for
type~II groups and explore under which conditions they can be imposed as a
symmetry. Since, as discussed in \Secref{sec:ProperCP}, we have not found any
higher--order class--inverting automorphism without a corresponding involutory
automorphism that has the same physical implications, we specialize to the case
of involutory automorphisms to simplify the discussion.

\subsubsection{Existence of CP transformations}

It has been shown in \cite{Damhus:1981} that matrices \UU[_i]{\rep{r}} which
solve \eqref{eq:consistency1} and hence allow for generalized CP transformations
\begin{equation}
 \rep[_i]{r}~\xmapsto{\CPgen}~ 
 \UU[_i]{\rep{r}}\,\rep[_i]{r}^*\;,
 \label{eq:constituent_trafo}
\end{equation}
exist for all irreducible representations $\rep r_i$ if and only if the group
exhibits a class--inverting automorphism. To simplify the discussion, one can
work in special bases that are particularly convenient for the analysis of CP
properties.\footnote{These bases can have other deficiencies, see
\Secref{sec:twoA}.} The general situation for class--inverting automorphisms is
discussed in \cite{Ecker:1987qp}; however, since we are dealing with
class--inverting and involutory automorphisms,  we know that the matrices
\UU[_i]{\rep{r}} are either symmetric or anti--symmetric,
$\UU[_i]{\rep{r}}^T=\pm \UU[_i]{\rep{r}}$ \cite{Damhus:1981}, cf.\ the
discussion around \Eqref{eq:U(anti)symmetric}.  This leads to even simpler
standard forms than in the general case. In fact, any unitary (anti--)symmetric
matrix $U$ can be written as
\begin{equation}
 U~=~W\,\Sigma\,W^T\;,
\end{equation}
with unitary $W$ and
\begin{equation}\label{eq:Sigma}
 \Sigma ~=~\left\{\begin{array}{ll}
 \Sigma_+~=~\mathbbm{1}\;, & \text{if $U$ is symmetric,}\\
 \Sigma_-~=~\left(\begin{array}{cc|c|cc}
    & 1 &    & &   \\
 -1 &   &    & &  \\
 \hline
 & & \ddots & &   \\
 \hline
 & &  & & 1\\
 & & & -1 &   \\
 \end{array}\right)
 \;, & \text{if $U$ is anti--symmetric.}
 \end{array}\right.
\end{equation}
Note that, since representation matrices always have full rank, the
anti--symmetric case does not arise for odd--dimensional irreps
\cite{Damhus:1981}, i.e.\ $\Sigma$ always has full rank. We can, hence, perform
the unitary basis change
\begin{equation}
  \rep[_i]{r}~\to~ \WW[_i]{r}^\dagger\,\rep[_i]{r}
  \;,\quad 
  \rhoR{_i}(g) ~\to~ \WW[_i]{\rep{r}}^\dagger\,\rhoR{_i}(g)\,\WW[_i]{\rep{r}} 
  \quad \forall~g \in \DiscreteGroup\;,
  \label{eq:CP_basis}
\end{equation}
such that in the new basis the matrices $\UU[_i]{\rep{r}}$ take the simple form
\begin{equation}
 \UU[_i]{\rep{r}} ~\to~ \WW[_i]{\rep{r}}^\dagger\,\UU[_i]{\rep{r}}\,\WW[_i]{\rep{r}}^*~=~\SIGMA[_i]{r}\;.
\end{equation}
For type II~A groups, all the $\SIGMA[_i]{r}$'s equal the identity matrix and the
new basis is a CP basis. In this basis all Clebsch--Gordan coefficients are real
\cite{Bickerstaff:1985jc}.

Let us now investigate under which conditions CP can be imposed as a
symmetry. In the most general case, we can write the contraction of two
multiplets $x$ and $y$ transforming in irreducible representations
$\rep{r}(x)=\rep[_x]{r}$ and  $\rep{r}(y)=\rep[_y]{r}$ to a representation
$\rep{r}(z)=\rep[_z]{r}$ as
\begin{equation}
 \left[\left(\myvec{x}\otimes\myvec{y}\right)_{\rep[_z]{r}}\right]_\mu 
 ~=~ 
 C_{\mu,\alpha\beta}\,x_\alpha\,y_\beta~=~x^T\,C_{\mu}\,y\;,
 \label{eq:contraction}
\end{equation}
where $\alpha$ and $\beta$ denote the vector indices of $\myvec{x}$ and
$\myvec{y}$, and $C_{\mu, \alpha\beta}$ are the Clebsch--Gordan coefficients for
the $\mu^\mathrm{th}$ component of the resulting representation vector. In the last step
we have switched to matrix notation, i.e.\ $C_\mu$ is a matrix, and $x$ and $y$
are vectors. We will also need the complex conjugate of the contraction, which
reads
\begin{equation}
 \left[\left(\myvec{x}\otimes\myvec{y}\right)^*_{\rep[_z]{r}}\right]_\mu 
 ~=~
 C^*_{\mu,\alpha\beta}\,x^*_\alpha\,y^*_\beta
 ~=~
 x^\dagger\,C^*_{\mu}\,y^*\;.
 \label{eq:contraction_conjugate}
\end{equation}
In what follows, we will refer to
$\left(\myvec{x}\otimes\myvec{y}\right)_{\rep[_z]{r}}$ as a ``meson'' and to $x$
and $y$ as its ``constituents''.

The generalized CP transformation acts on $x$ and $y$ as specified in
\Eqref{eq:constituent_trafo} with the matrices 
$(\UU[_x]{\rep{r}})_{\alpha\beta}$ and $(\UU[_y]{\rep{r}})_{\alpha\beta}$, 
respectively. From this, we can derive the CP transformation of the meson
$\left(\myvec{x}\otimes\myvec{y}\right)_{\rep[_z]{\myvec{r}}}$,
\begin{equation}
 \left[\left(\myvec{x}\otimes\myvec{y}\right)_{\rep[_z]{r}}\right]_\mu
 ~=~
 x^T\,C_{\mu}\,y
 ~\xmapsto[(\ref{eq:constituent_trafo})]{\CPgen}~ 
 x^\dagger\,\UU[_x]{r}^T\,C_{\mu}\,\UU[_y]{r}\,y^*\;.
 \label{eq:meson_trafo}
\end{equation}
In general, also a multiplet $z$ in the representation $\rep[_z]{r}$  will
transform under the generalized CP transformation with some matrix
$\UU[_z]{\rep{r}}$,  such that one might demand that
\begin{equation}
 \left[\left(\myvec{x}\otimes\myvec{y}\right)_{\rep[_z]{r}}\right]_\mu
 ~\xmapsto{\CPgen}~ 
 (\UU[_z]{\rep{r}})_{\mu\nu}\,\left[\left(\myvec{x}\otimes\myvec{y}\right)^*_{\rep[_z]{r}}\right]_{\nu}~
 \stackrel{\eqref{eq:contraction_conjugate}}{=}~(\UU[_z]{\rep{r}})_{\mu\nu}\,\left[x^\dagger\,{C_{\nu}}^*\,y^*\right]\;.
 \label{eq:bare_rep_trafo}
\end{equation}
Comparing \eqref{eq:bare_rep_trafo} with \eqref{eq:meson_trafo}, we obtain the condition
\begin{equation}
 \UU[_x]{r}^T\,C_{\mu}\,\UU[_y]{r}~\stackrel{!}{=}~(\UU[_z]{r})_{\mu\nu}\,{C_{\nu}}^*\;,
\label{eq:consistency_condition}
\end{equation}
for the consistency of meson and constituent transformations. Recall that the
matrices $\UU[_x]{r}$, $\UU[_y]{r}$, and $\UU[_z]{r}$ are representations of a
class--inverting automorphism and hence  are given by the solutions of
\eqref{eq:consistency1}. However, the fact that the matrices fulfill
\eqref{eq:consistency1} does not imply that they also solve
\eqref{eq:consistency_condition}. In other words $\UU[_x]{r}$, $\UU[_y]{r}$, and
$\UU[_z]{r}$, in general, do not satisfy \eqref{eq:consistency_condition}.
The existence of an automorphism for which the matrices which solve
\eqref{eq:consistency1} also satisfy \eqref{eq:consistency_condition} is a
non--trivial property of a group.\footnote{The Clebsch--Gordan coefficients
determine a group up to isomorphism \cite{Feit:1967}.}

At this point, let us note that we are free to re--define phases in the
definition of
\begin{itemize}
 \item the Clebsch--Gordan coefficients $C_{\mu,\alpha\beta}$, i.e.\ one global
  phase for each $\rep[_z]{r}$ appearing in the contraction of $\rep[_x]{r}$ and
  $\rep[_y]{r}$;
 \item the CP transformations \UU[_x]{\rep{r}}, \UU[_y]{\rep{r}} and
  \UU[_z]{\rep{r}}.
\end{itemize}
This freedom of global phase choices can obscure possible solutions to
\eqref{eq:consistency_condition} (cf.\ the discussion around
\eqref{eq:HLSdoublettrafo}).

Whether or not one can solve \eqref{eq:consistency_condition} is most
conveniently analyzed in the basis \eqref{eq:CP_basis}, in which
\eqref{eq:consistency_condition} reads
\begin{equation}\label{eq:consistency_condition_prime_basis}
 \SIGMA[_x]{r}^T\,C'_{\mu}\,\SIGMA[_y]{r}
 ~\stackrel{!}{=}~
 (\SIGMA[_z]{r})_{\mu\nu}\,{(C'_{\nu})}^*\;.
\end{equation}
Here we have introduced the basis--transformed Clebsch--Gordan coefficients
\begin{equation}
 C'_\mu~:=~\WW[_x]{r}^T\,C_\mu\,\WW[_y]{r}\;.
\end{equation}

Whether or not \eqref{eq:consistency_condition} (or equivalently
\eqref{eq:consistency_condition_prime_basis}) can be solved depends on the
specific automorphism we use to define CP and, hence, on whether the group is
type II~A or type II~B. In the case of type II~A groups, the underlying
automorphism of the CP transformation is a BDA and, hence, all \SIGMA[_i]{r}'s
equal the identity (because all matrices \UU[_i]{\rep{r}} are unitary and
symmetric). Therefore, all the Clebsch--Gordan coefficients are real
\cite{Bickerstaff:1985jc} such that \Eqref{eq:consistency_condition_prime_basis}
is trivially fulfilled. This statement is trivial in the CP basis
but, of course, holds for all other bases as well. Hence the \UU[_i]{\rep{r}}
indeed provide us with a solution to \Eqref{eq:consistency_condition}. In other
words, for type II~A groups one can always find matrices \UU[_i]{\rep{r}} such
that the transformation of a meson under \CPgen\ follows from the (generalized)
CP transformation properties of its constituents. We remark that, as we shall
demonstrate in an example in \Secref{sec:twoA}, the CP basis often turns out not
to be the most convenient choice for analyzing a model.

If instead the class--inverting and involutory automorphism used to define CP is
not a BDA, as is always the case for type~II~B groups, some of the
\UU[_i]{\rep{r}}'s are anti--symmetric  and the existence of a solution to
\eqref{eq:consistency_condition_prime_basis} is not guaranteed. One can,
however, use the symmetry properties of \SIGMA[_x]{r}, \SIGMA[_y]{r} and
\SIGMA[_z]{r} to check whether a solution is possible. If both \SIGMA[_x]{r} and
\SIGMA[_y]{r} are either symmetric or anti--symmetric, \SIGMA[_z]{r} has to be
symmetric, while in the mixed case \SIGMA[_z]{r} has to be anti--symmetric. In
all other cases, \eqref{eq:consistency_condition_prime_basis} has no solution.
In order to see this, consider, for example, the case of two representations
$\rep[_x]{r}$ and $\rep[_y]{r}$ transforming with two symmetric matrices
\SIGMA[_x]{r} and \SIGMA[_y]{r}, and contracting to a representation
$\rep[_z]{r}$ transforming with an anti--symmetric  $\SIGMA[_z]{r}$. Then we see
that \eqref{eq:consistency_condition_prime_basis} implies that
\begin{equation}\label{eq:CGcontradiction}
 C'_1~=~(C'_2)^*
 \quad\text{and}\quad
 C'_2~=~-(C'_1)^*\;,
\end{equation}
such that Clebsch--Gordan coefficients $C_\mu'$ have to vanish, which is
obviously a contradiction. This means that a solution to
\eqref{eq:consistency_condition_prime_basis} is not possible.  Hence, if a group
allows for such mixed contractions then it is not possible to make all mesons
transform in consistency with their constituents. This has striking consequences
for the CP properties of a model because, as we will also show in an
explicit example in \Secref{sec:twoB}, physical CP conservation then implies
the absence of the problematic terms from the Lagrangean.

\subsubsection{CP transformations and constraints on couplings}

Let us now discuss how the CP transformation \eqref{eq:constituent_trafo}
constrains the physical coupling coefficients of a model. Consider a model with
some fields furnishing the representations $\rep[_x]{r}$ and $\rep[_y]{r}$. 
The presence of a
contraction \eqref{eq:contraction} (with coupling coefficient $c$) in the
Lagrangean implies also the presence of the conjugate contraction
\eqref{eq:contraction_conjugate} (with the complex conjugate coupling) in order
to guarantee the reality of the Lagrangean.  The couplings are, up to the global
factor $c$, given by the Clebsch--Gordan coefficients $C_\mu$. If a theory is
invariant under the symmetry, all contractions have to be trivial singlets and
therefore  the only relevant case is when $(\UU[_z]{r})_{\mu\nu}$ is trivial and
$\mu$ only takes on the value 1. The condition for the term $c\,x^TC_{\mu}y$ to
conserve CP then is given by
\begin{equation}
 c\,\UU[_x]{r}^T\,C_{\mu=1}\,\UU[_y]{r}~\stackrel{!}{=}~c^*\,{C^*_{\mu=1}}\;,
\label{eq:CP_consistency_condition}
\end{equation}
which is a simplified version of the consistency condition
\eqref{eq:consistency_condition}. If the corresponding conditions are fulfilled
for all contractions present in the Lagrangean, CP is conserved. Note that
adding a phase to the generalized CP transformations \UU[_x]{r} or \UU[_y]{r} is
nothing but a simple rephasing of fields.

We conclude that, for type II A groups, CP is automatically conserved if there
is enough rephasing freedom of fields to render all couplings real, i.e.\ the
corresponding phases unphysical, simply because equations
\eqref{eq:CP_consistency_condition} are already fulfilled from the group
structure. A sufficient number of field redefinitions, however, may not be
possible in generic models  (see e.g.\ \cite{Haber:2012np} for criteria), in
which case CP can be explicitly violated by the physical phases of couplings.
Turning this around, we see that imposing the generalized CP transformation as a
symmetry for type II~A groups forces all couplings to be real (up to the
above--mentioned freedom coming from field redefinitions). The situation for
type II~A groups, therefore, is somewhat similar to the familiar case where only
continuous symmetries such as \SU{N} are present.

On the other hand, in the case of type~II~B groups, some representations may
contract in such a way as to make it impossible to solve the appropriate
analogue of \eqref{eq:CP_consistency_condition}. Thus, the corresponding terms
cannot be part of the Lagrangean if CP is to be conserved. It is clear from the
discussion in \Secref{sec:ProperCP} that imposing CP in this case implies the 
presence of an additional $\Z2$ symmetry related to $V$. It is this  $\Z2$
which prohibits exactly the problematic contractions. That is, type II~B groups
can have the unusual property that CP invariance forbids certain couplings
rather than just restricting the phases of the coefficients. 

Let us remark that, in principle, it is conceivable that the structure of a type
II~B group is such that it does not allow for CP violating contractions. This is
the case if and only if the \Z2 symmetry related to the element represented by
$V$ is already part of the group. For all examples we have found and given in
\Tabref{tab:typeIIB} this is not the case, and the groups have to be extended
(trivially) by the \Z2 in order to warrant CP conservation. Although we cannot
make a general statement on when a group is extended, we can prove it for the
special case of ambivalent type~II~B groups and inner automorphisms.\footnote{A
group is called ambivalent if it possesses only real and pseudo--real
irreducible representations. For such groups, inner automorphisms are
class--inverting.} Consider first the identity automorphism $u(g)=g$. For this
automorphism, the twisted Frobenius--Schur indicator coincides with the
original, un--twisted indicator, i.e.\
$\mathrm{FS}_{u=\mathrm{id}}(\rep[_i]{r})=\pm1$ for real and pseudo--real irreps
$\rep[_i]{r}$, respectively. Hence, $V_{\rep[_i]{r}}=\mathbbm{1}$ for real and
$V_{\rep[_i]{r}}=-\mathbbm{1}$ for pseudo--real representations. However, it has
been shown in \cite{Damhus:1981} that an ambivalent group with an element whose
representation matrices are given by these particular $V_{\rep[_i]{r}}$ has a
basis with real CG coefficients. Thus, \DiscreteGroup\ would be of type~II~A,
which contradicts the assumption that the group is type~II~B. Hence, there can
be no such group element. The argument can be extended to all inner
automorphisms noting that the $V_{\rep[_i]{r}}$ belonging to these automorphisms
are connected to the $V_{\rep[_i]{r}}$ of the identity by multiplication with a
group element (cf.\ \Appref{app:higherOrder} and footnote~\ref{fn:Vinner}). In
summary, imposing a CP transformation which is defined via in inner automorphism
of an ambivalent type~II~B group always extends the finite symmetry (trivially)
by a \Z2 factor.

We conclude that for groups of type II it is always possible to
define a physical CP transformation in a model--independent way. Whether or not
it is broken then depends on the details of the model. Below, in
sections~\ref{sec:twoA} and \ref{sec:twoB} we will present examples
illustrating the CP properties of such groups.

\section{Examples}
\label{sec:Examples}

\subsection{Example for a type~I group: \texorpdfstring{$\boldsymbol{\Delta(27)}$}{Delta(27)}}
\label{sec:Delta27}

In what follows, we substantiate the statement that type~I groups generically
violate CP, focusing on a toy model based on the group $\Delta(27)$.

\subsubsection{Decay amplitudes in a toy example based on
\texorpdfstring{$\boldsymbol{\Delta(27)}$}{Delta(27)}}
\label{sec:Delta27decay}

Let us consider a toy model based on the symmetry group $\Delta(27)$. The
necessary details on the group are summarized in \Appref{app:Delta27}. We
introduce three complex scalars $S$, $X$ and $Y$ transforming as $\rep[_0]{1}$,
$\rep[_1]{1}$ and $\rep[_3]{1}$ as well as two sets of fermions $\Psi$ and
$\Sigma$, transforming as $\rep{3}$ each. Furthermore, we assume that there is a
\U1 symmetry under which $Y$ is neutral, $\Psi$ has charge $q_\Psi$, $\Sigma$
has charge $q_\Sigma$, and $S$ and $X$ both have charge
$q_X=q_\Psi-q_\Sigma\neq0$. Then the renormalizable interaction Lagrangean
reads\footnote{There might also a cubic $Y$ coupling, which is, however,
irrelevant for our discussion.}
\begin{align}
 \mathscr{L}_\mathrm{toy}
 ~=~&
 f\,\left[
        S_{\rep[_0]{1}}\otimes \left(\overline{\Psi}\,\otimes\,\Sigma\right)_{\rep[_0]{1}}
 \right]_{\rep[_0]{1}}
 +g\,\left[
        X_{\rep[_1]{1}}\otimes \left(\overline{\Psi}\,\otimes\,\Sigma\right)_{\rep[_2]{1}}
 \right]_{\rep[_0]{1}}
 \nonumber\\
 &
 +h_\Psi\,\left[
        Y_{\rep[_3]{1}}\otimes \left(\overline{\Psi}\,\otimes\,\Psi\right)_{\rep[_6]{1}}
 \right]_{\rep[_0]{1}}
 +h_\Sigma\,\left[
        Y_{\rep[_3]{1}}\otimes \left(\overline{\Sigma}\,\otimes\,\Sigma\right)_{\rep[_6]{1}}
 \right]_{\rep[_0]{1}}
 +\text{h.c.}\nonumber\\
 ~=~&
 F^{ij}\,S\,\overline{\Psi}_i\Sigma_j
 +G^{ij}\,X\,\overline{\Psi}_i\Sigma_j
 +H_\Psi^{ij}\,Y\,\overline{\Psi}_i\Psi_j
 +H_\Sigma^{ij}\,Y\,\overline{\Sigma}_i\Sigma_j
 +\text{h.c.}\;.\label{eq:Ltoy}
\end{align} 
The ``Yukawa'' matrices are given by
\begin{equation}
 F~=~f\,\mathbbm{1}_3\;,\quad
 G ~=~ g\,\left(
\begin{array}{ccc}
 0 & 1 & 0 \\
 0 & 0 & 1 \\
 1 & 0 & 0 \\
\end{array}
\right)
\quad\text{and}\quad
 H_{\Psi/\Sigma} ~=~ h_{\Psi/\Sigma}\,\left(
\begin{array}{ccc}
 1 & 0 & 0 \\
 0 & \omega^2 & 0 \\
 0 & 0 & \omega \\
\end{array}
\right)\;,
\end{equation}
where $f$, $g$, $h_{\Psi}$, and $h_{\Sigma}$ are complex couplings and we define
$\omega:=\mathrm{e}^{2\pi\,\I/3}$.

\begin{figure}[t!]
\centerline{
\begin{tabular}{cc}
\subfigure[\label{fig:YdecayTree1}]{%
\includegraphics{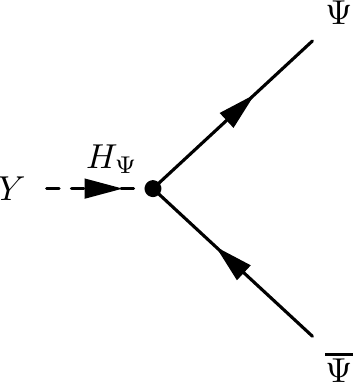}}
\quad & \quad
\subfigure[\label{fig:Ydecay1Loop4}]{%
\includegraphics{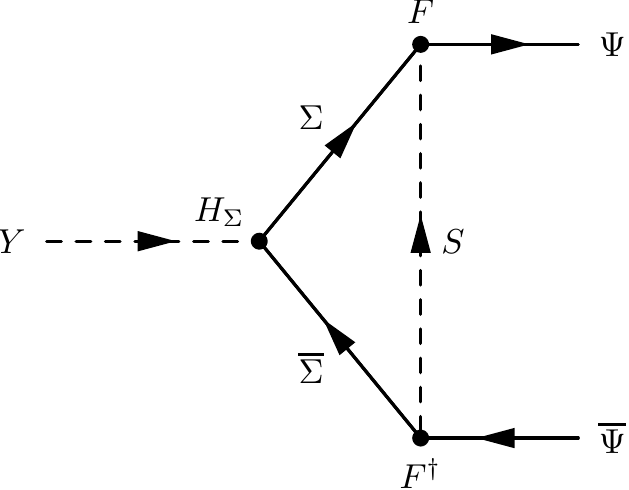}}
\\
\subfigure[\label{fig:Ydecay1Loop1}]{%
\includegraphics{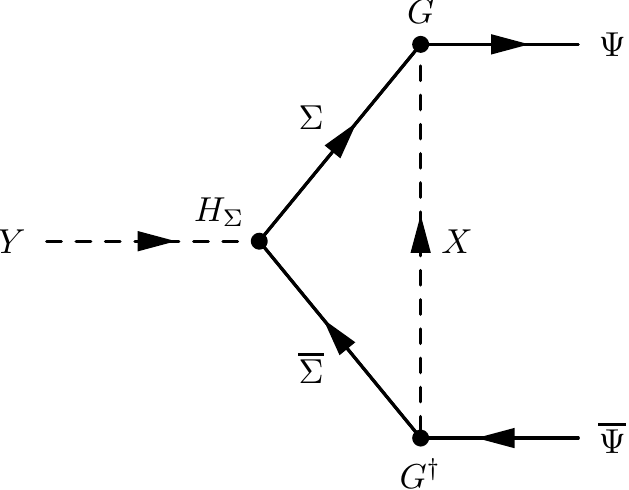}}
\quad & \quad
\subfigure[\label{fig:Ydecay1Loop2}]{%
\includegraphics{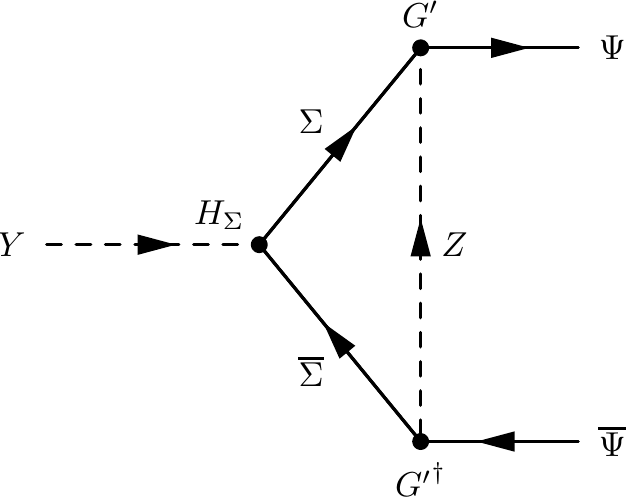}}
\end{tabular}}
\caption{Diagrams relevant for CP violation in
$Y\to\bar{\Psi}\Psi$ at tree level and 1--loop.}
\label{fig:DecayAsymmetry}
\end{figure}

Let us now study the decay $Y\to\overline{\Psi}\Psi$. Interference between
tree--level and one--loop diagrams (figures~\ref{fig:YdecayTree1}--
\ref{fig:Ydecay1Loop1}) leads to a CP asymmetry
$\varepsilon_{Y\to\overline{\Psi}\Psi}$, which is proportional to
\begin{align}
 \varepsilon_{Y\to\overline{\Psi}\Psi}
 &~\propto~
 \im\left[I_S\right]\,\im\left[\tr\left(F^\dagger\,H_\Psi\,F\,H_\Sigma^\dagger\right)\right]
 +
 \im\left[I_X\right]\,\im\left[\tr\left(G^\dagger\,H_\Psi\,G\,H_\Sigma^\dagger\right)\right]
 \nonumber\\
 &
 ~=~
 |f|^2\,\im\left[I_S\right]\,\im\left[h_\Psi\,h_\Sigma^*\right]+
 |g|^2\,\im\left[I_X\right]\,\im\left[\omega\,h_\Psi\,h_\Sigma^*\right]\;.
 \label{eq:DecayAsymmetry}
\end{align}
Here $I_{S}=I(M_S,M_Y)$ and $I_{X}=I(M_X,M_Y)$ denote appropriate phase space
factors and the loop integral, which are non--trivial functions of the masses of
$S$ and $Y$, and $X$ and $Y$, respectively. Note that
$\varepsilon_{Y\to\overline{\Psi}\Psi}$ is 
\begin{itemize}
\item[(i)] invariant under rephasing of the fields,
\item[(ii)] independent of the phases of $f$ and $g$, and
\item[(iii)] independent of the chosen basis as it is proportional to the trace
of coupling matrices.
\end{itemize}
Notice, however,
that the asymmetry can vanish if there is a cancellation between the two terms,
which would require a delicate adjustment of the relative phase
$\varphi:=\arg(h_\Psi\,h_\Sigma^*)$ of $h_\Psi$ and $h_\Sigma$.  In what
follows, we will argue that if such a cancellation occurs, this is either (i) a
consequence of a larger discrete symmetry than $\Delta(27)$ being present or
(ii) it is not immune to quantum corrections.

In the first case, a new symmetry has to be present which relates $S$ and $X$ in
such a way as to guarantee $M_S=M_X$ and $|g|=|f|$, as well as $h_\Psi$ and
$h_\Sigma$ to warrant $\varphi=-2\pi/6$. Clearly, this cannot be due to an outer
automorphism and, hence, no CP transformation of a $\Delta(27)$ setup since such
transformations never relate the trivial singlet $\rep[_0]{1}$ to other
representations. If such a symmetry exists, it has to enhance the original
flavor symmetry of the setup, and it is, therefore, no longer appropriate to
speak of a $\Delta(27)$ model.

In the second case, given that  $\im\left[I_S\right]\ne\im\left[I_X\right]$ for
$M_S\ne M_X$, an adjustment which cancels the asymmetry will require
$\arg(h_\Psi\,h_\Sigma^*)$ to be different from $-2\pi/6$ in general. Note that
the diagrams of figures~\ref{fig:Ydecay1Loop4} and \ref{fig:Ydecay1Loop1} also
yield vertex corrections which are relevant for the renormalization group
equations (RGEs) for $h_\Psi$ and $h_\Sigma$. These equations are given
by\footnote{Note that $G\,H_{\Psi/\Sigma}\,G^{\dagger}=\omega^2\,H_{\Psi/\Sigma}$.}
\begin{subequations}
\begin{align}
 16\pi^2\,\frac{\D h_\Psi}{\D t}
 &~=~ h_\Psi\, \left(a\,|h_\Psi|^2+b\,|h_\Sigma|^2+ \dots\right)
 +c\,h_\Sigma\,\left[|f|^2+\omega^2\,|g|^2\right]\;,
 \\
 16\pi^2\,\frac{\D h_\Sigma}{\D t}
 &~=~ h_\Sigma\, \left(a\,|h_\Sigma|^2+b\,|h_\Psi|^2+ \dots\right)
 +c\,h_\Psi\,\left[|f|^2+\omega\,|g|^2\right]\;,
\end{align}
\end{subequations}
where $t=\ln(\mu/\mu_0)$ is the logarithm of the renormalization scale, $a$, $b$
and $c$ are real coefficients, and the omission represents terms like the square
of the gauge coupling. This leads to an RGE for $h_\Psi\,h_\Sigma^*$ with the
structure
\begin{equation}
 16\pi^2\,\frac{\D }{\D t}\left(h_\Psi\,h_\Sigma^*\right)
 ~=~h_\Psi\,h_\Sigma^*\times \text{real}+c\,\left(|h_\Psi|^2+|h_\Sigma|^2\right)\,
 \left[|f|^2+\omega^2\,|g|^2\right]\;.
\end{equation}
The only value of the relative phase $\varphi$ that is stable under the RGE
is, thus, given by
$\varphi=\varphi_*=\arg\left(|f|^2+\omega^2\,|g|^2\right)$. Therefore, we see
that, if one imposes that \eqref{eq:DecayAsymmetry} vanishes at one
renormalization scale, this relation will be violated at other scales provided
that $\im[I_S]\ne\im[I_X]$. Hence, even if one adjusts the phases of $h_\Psi$
and $h_\Sigma$ by hand, this relation will be destroyed by quantum corrections.
We note that such considerations can always be used in order to see if a
particular relation is a consequence of a symmetry, in which case it has to
respect quantum corrections, or not.

Altogether, we conclude that this simple setting based on $\Delta(27)$
generically violates CP.  The reason behind this is that any conceivable
generalized CP transformation  is inconsistent with $\Delta(27)$ --- simply due
to the fact that the group (with the field content chosen) does not allow for a
class--inverting automorphism. Hence any transformation which relates
each field to its complex conjugate, if simply imposed on the theory, would map
$\Delta(27)$--invariants to non--invariants, similarly to what happened
around \Eqref{eq:canonicalCP}, but without the possibility of ``repairing'' the
transformation.

Let us point out that due to the peculiarities of the example model, there is
neither a \U1 charge asymmetry nor a left--right asymmetry produced by the CP
violating decay. In general, however, there seems to be no obstacle in
constructing such models. Note also that it is, in principle, possible
to distinguish $Y$ from $Y^*$ by measuring the relative branching fraction of
the decays to  $\bar{\Psi}\Psi$ and $\bar{\Sigma}\Sigma$, where a pure sample
of, say, $Y$ particles could be generated if we couple it to a pair of chiral
fermions.

\subsubsection{Spontaneous CP violation with calculable CP phases}

Let us next discuss how one could possibly restore CP invariance by enlarging
the symmetry group, and how this can lead to the possibility of breaking CP
spontaneously with calculable CP phases. Imposing a symmetry that ensures the
vanishing of the decay asymmetries is possible if the field content of a theory
allows us to combine fields which transform as irreducible representations of
$\Delta(27)$ to multiplets of a larger group, which then itself has an
appropriate class--inverting involutory automorphism, i.e.\ a proper CP
transformation. In order to see how this could work, let us replace $S$ by a
field $Z$, transforming in the non--trivial one--dimensional representation
$\rep[_8]{1}$ and still carrying the same \U1 charge as $X$. This will lead to
an allowed coupling
\begin{equation}
 \mathscr{L}_\mathrm{toy}^Z~=~
 g'\,\left[
        Z_{\rep[_8]{1}}\otimes \left(\overline{\Psi}\,\otimes\,\Sigma\right)_{\rep[_4]{1}}
 \right]_{\rep[_0]{1}}+\text{h.c.}
 ~=~
 (G')^{ij}\,Z\,\overline{\Psi}_i\Sigma_j+\text{h.c.}
\end{equation}
with the ``Yukawa'' matrix
\begin{equation}
 G'~=~g'\,
\left(
\begin{array}{ccc}
 0 & 0 & \omega^2 \\
 1 & 0 & 0  \\
 0 & \omega & 0
\end{array}
\right)\;.
\end{equation}
Instead of the process shown in \Figref{fig:Ydecay1Loop4}, one now has to take
into account the one--loop diagram for the $Y$ decay displayed in
\Figref{fig:Ydecay1Loop2}, which contributes to the decay asymmetry as
\begin{equation}
 \varepsilon_{Y\to\overline{\Psi}\Psi}^Z
 ~\propto~|g'|^2\,\im\left[I_Z\right]\,\im[\omega^2\,h_\Psi\,h_\Sigma^*]\;.
\end{equation}
The statement that CP is generically violated still holds. However, the total CP
asymmetry of the $Y$ decay vanishes if (i) $M_Z=M_X$, (ii) $|g|=|g'|$, and (iii)
$\varphi=0$. It is possible to understand this CP conserving spot in parameter space
from the fact that one can enhance the flavor symmetry beyond $\Delta(27)$. Here
this can happen via the (outer) automorphism ${u_3}$ of $\Delta(27)$ (details
are given in \Appref{app:Delta27}) which transforms
\begin{equation}
 X~\xleftrightarrow{~u_3~}~Z\;,\quad Y~\xrightarrow{~u_3~}~ Y\;,\quad 
 \Psi~\xrightarrow{~u_3~}~ U_{u_3}\Sigma^\ChargeC\quad\text{and}\quad
 \Sigma~\xrightarrow{~u_3~}~ U_{u_3}\,\Psi^\ChargeC\;,
\end{equation}
with $U_{{u_3}}$ given in \Eqref{eq:UDelta27}. This symmetry is consistent with
the \U1 symmetry (for the choice $q_\Sigma=-q_\Psi$) and naturally ensures
relations (i)--(iii), thereby granting the absence of CP violation. Let us
stress that this is not a CP symmetry of the $\Delta(27)$ model, but instead
enhances the flavor symmetry of the setup from $\Delta(27)$ to
$\text{SG}(54,5)$ --- and this bigger group itself then has an appropriate
class--inverting involutory automorphism which ensures CP
conservation. The bigger symmetry can be constructed as the semi--direct product
of $\Delta(27)$  and the symmetry ${u_3}$
\cite{Holthausen:2012dk,Feruglio:2012cw}. Under this bigger symmetry, the
previously distinct fields $X$ and $Z$ get combined to a doublet, $\Psi$ and
$\Sigma^\ChargeC$ get combined to a hexaplet and $Y$ stays in a non--trivial
one--dimensional representation. Then, since we have enough fields at hand to
render all coupling phases unphysical, the class--inverting involutory
automorphism of $\text{SG}(54,5)$, which is a physical CP transformation, is an
accidental symmetry of the setting.

Note that if the relations (i)--(iii) are fulfilled, the quantum
corrections to the relative phases of $h_\Psi$ and $h_\Sigma$ vanish. 
This substantiates the statement that one can use the behavior under the
renormalization group to check whether or not certain
relations are caused by a symmetry.

Interestingly, it is possible to spontaneously break the larger group
$\text{SG}(54,5)$ down to $\Delta(27)$ by the VEV of a (\U1 neutral) field
$\phi$ in the real non--trivial one--dimensional representation of
$\text{SG}(54,5)$. The field $\phi$ couples to the scalars, thus giving rise to
a mass splitting,
\begin{equation}
  \mathscr{L}_\mathrm{toy}^\phi ~\supset~ M^2\,\left(|X|^2 + |Z|^2\right) 
  + \left[\frac{\mu}{\sqrt{2}}\,\langle \phi \rangle \, \left(|X|^2 - |Z|^2\right)
  +\text{h.c.}\right]\;,
\end{equation}
with $\mu$ denoting a mass parameter. However, $\phi$ does not couple to, i.e.\
alter, the Yukawa couplings of $X$, $Y$, and $Z$ at the renormalizable level.
Note that, given an appropriate coupling, the $\phi$ VEV will also split the
masses of the fermions, thus making them distinguishable.

Therefore, after the breaking, relations (ii) and (iii), i.e.\ the equalities
$|g|=|g'|$ and $h_{\Psi}=h_{\Sigma}$, still hold, while due to the mass
splitting the relation (i) gets destroyed, i.e.\ $M_X\neq M_Z$.
As a consequence, CP is violated spontaneously and all the phases which
appear in the CP asymmetry
\begin{equation}
 \varepsilon_{Y\to\overline{\Psi}\Psi}~\propto~
 |g|^2\,|h_\Psi|^2\,\im\left[\omega\right]\,\left(\im\left[I_X\right]-\im\left[I_Z\right]\right)\;,
 \label{eq:XZDecayAsymmetry}
\end{equation}
are independent of the couplings, i.e.\ calculable.

We have, hence, obtained a simple recipe for constructing models of
spontaneous CP breaking. One starts with a type II group
$\DiscreteGroup_\mathrm{II}$ which contains (and can be spontaneously broken
down to) a type I group $\DiscreteGroup_\mathrm{I}$. At the level of
$\DiscreteGroup_\mathrm{II}$, one imposes the generalized CP transformation,
such that at this level CP is conserved. After the spontaneous breaking
$\DiscreteGroup_\mathrm{II}\to\DiscreteGroup_\mathrm{I}$ CP will, at least
generically, be broken. In the example discussed above, the CP phases are even
calculable. A more detailed discussion of these issues will be presented in a
subsequent publication.

\subsubsection{CP--like symmetries}
\label{sec:CPlike}
 
Let us also emphasize that not every outer automorphism which is imposed as a
symmetry does lead to physical CP conservation.  Consider for example the outer
automorphism ${u_5}$ given in \eqref{eq:u_5}, which
is the same as $u$ in \cite{Holthausen:2012dk}.\footnote{In \HLS\ the
automorphism is defined as  $u:~(A,B)\to(A\, B^2\, A\, B , A\, B^2\, A^2)$ while
in \eqref{eq:u_5}  $u_5:~ (A,B)\to (B\, A^2\, B^2, A\, B^2\, A^2 )$. However,
as  $A\, B^2\, A = B\, A^2\, B$, these operations coincide.}  The only way in
which ${u_5}$ can act consistently with all symmetries (again for the choice
$q_\Sigma=-q_\Psi$) is to exchange the fermions,
\begin{equation}\label{eq:Delta27additionalTrafo}
 X~\xrightarrow{~u_5~}~X^*\,,\quad Z~\xrightarrow{~u_5~}~Z^*\,,\quad Y~\xrightarrow{~u_5~} ~Y^*\,,\quad 
 \Psi~\xrightarrow{~u_5~}~ U_{{u_5}}\,\Sigma\,,\quad\text{and}\quad
 \Sigma~\xrightarrow{~u_5~}~ U_{{u_5}}\,\Psi\;.
\end{equation}
Clearly, this transformation maps fields with opposite \U1 charges onto each
other, i.e.\ acts like a charge conjugation on the \U1. Hence  this is not an
enhancement of the flavor symmetry. It is, however, not a physical CP symmetry
either since not all representations of $\Delta(27)$  are mapped
to their complex conjugate representations, in particular $\rep{3}\to\rep{3}$.
That is, if one were to entertain the possibility that, in a more sophisticated
model, the \rep{3}--plet describes some fields that are connected to the
standard model, this transformation would not entail a physical CP
transformation (see also our earlier discussion in \Secref{sec:ProperCP}).  Note
that imposing \eqref{eq:Delta27additionalTrafo} will enforce equality between
the decay amplitudes of $Y\to\overline{\Psi}\Psi$ and
$Y^*\to\overline{\Sigma}\Sigma$ but none of the relations (i)--(iii) is
fulfilled and thus the physical CP asymmetry of the $Y$ decay, 
$\varepsilon_{Y\to\overline{\Psi}\Psi}$, is still non--vanishing, i.e.\ physical
CP is still violated. For these reasons we prefer to call such a symmetry a
``CP--like symmetry'' (see \Secref{sec:ProperCP}). In particular, we disagree
with the statement made by \HLS\ that an arbitrary outer automorphism can serve
as a physical CP transformation.

To conclude the discussion of the example, we emphasize that the CP violation in
$\Delta(27)$ exists solely due to the properties of the symmetry group and is
independent of any arguments based on spontaneously breaking this or other
symmetries. Yet, as we have seen,  it is possible to have settings in which
a bigger, CP conserving type II symmetry gets spontaneously broken down to
$\Delta(27)$. In this case, we have found that the physical CP violating phases
are predicted by group theory.

\subsubsection{CP conservation in models based on \texorpdfstring{$\boldsymbol{\Delta(27)}$}{Delta27}}

The alert reader may now wonder how it is possible that CP gets broken
spontaneously in $\Delta(27)$--based models \cite{Branco:1983tn}, i.e.\ how can
it be that there is CP conservation to start with. This is because in the model
discussed in \cite{Branco:1983tn} only triplet representations are introduced,
and there exist involutory automorphisms of $\Delta(27)$ for which the \FSI's
for the triplets equal 1 (see \Tabref{tab:Delta27FSI} in \Appref{app:Delta27},
and \cite{Nishi:2013jqa} for examples). This allows one to impose a consistent
CP transformation for this non--generic setting. However, once one amends the
setting by more than two non--trivial one--dimensional representations, this
will no longer be possible.

In summary, we see that models based on $\Delta(27)$ generically violate CP.
This can be avoided by
\begin{enumerate}
 \item increasing the (flavor) symmetry beyond $\Delta(27)$;
 \item considering settings in which only a special subset of representations is
 introduced.
\end{enumerate}

\subsubsection{Comments on the possible origin of a \texorpdfstring{$\boldsymbol{\Delta(27)}$}{Delta(27)} symmetry}

It may also be interesting to see how CP violation by discrete flavor symmetries
originates from some ``microscopic'' theory.  In \cite{Merle:2011vy} it was
studied how \SU3 can be broken to $\Delta(27)$, yet the discussion is based on
certain invariants and it is not clear if or how one can achieve this breaking
by giving VEVs to certain representation (cf.\ the discussion in
\cite{Adulpravitchai:2009kd}). However, given that $\Delta(27)$ does not allow
for a proper CP transformation while \SU3 does, it is tempting to speculate that
such a breaking, if possible, will also break CP spontaneously.

In \cite{Kobayashi:2006wq} it was shown how non--Abelian discrete flavor
symmetries arise in certain orbifold compactifications but no type I groups were
found. A $\Delta(27)$ symmetry can arise from space--group rules  in
non--Abelian heterotic orbifolds
\cite{Fischer:2012qj,Fischer:2013qza}\footnote{We thank P.~Vaudrevange for
pointing this out.} but it is not yet clear what the (massless) matter content
of such settings is, i.e.\ if there are representations with vanishing \FSI\ 
for all involutory automorphisms. Similar comments apply to
\cite{Berasaluce-Gonzalez:2013bba}, where, in a local construction, also a
$\Delta(27)$ symmetry was found.

\subsection{Example for a type~II~A group: \texorpdfstring{$\boldsymbol{\Tprime}$}{T'}}
\label{sec:twoA}

The group \Tprime, which is the double covering group of $\mathrm{A}_4$, is an
example for a group that admits a basis with real Clebsch--Gordan coefficients.
Information on the group structure of \Tprime\ and the tensor product
contractions in different bases can be found in \Appref{app:Tprime}.

\Tprime\ has a unique outer automorphism, which swaps each representation
with its complex conjugate representation, i.e.\ which is class--inverting. One
particular, involutory representative\footnote{By definition, all other
possible choices of automorphisms representing the unique outer automorphism of
\Tprime\ are connected to our choice by inner automorphisms.} of this outer
automorphism is given by
\begin{align}\label{eq:TprimeAutomorphism}
 u & ~:~ (S,T)\,\rightarrow\,(S^3, T^2)       
 & &~\curvearrowright~ 
 \rep[_i]{1}~\rightarrow~\UU[_i]{\rep{1}}\,\rep[_i]{1}^*\;,
 \quad
 \rep[_i]{2}~\rightarrow~\UU[_i]{\rep{2}}\,\rep[_i]{2}^*\;,
 \quad
 \rep{3}~\rightarrow~\UU{\rep{3}}\,\rep{3}^*\;.
\end{align}
One can confirm that this automorphism is
indeed a Bickerstaff--Damhus automorphism from the twisted Frobenius--Schur
indicators in \Tabref{tab:TPrimeFSI}.
\begin{table}[t]
\centering
\begin{tabular}{|c|ccccccc|}
\hline
 $\rep R$ & \rep[_0]{1} & \rep[_1]{1} & \rep[_2]{1} & \rep[_0]{2} & \rep[_1]{2} & \rep[_2]{2} & \rep{3} \\
\hline
$\TFS{u}(\rep R)$ & 1 & 1 & 1 & 1 & 1 & 1 & 1 \\
\hline
\end{tabular}
\caption{Twisted Frobenius--Schur indicators for the automorphism
\eqref{eq:TprimeAutomorphism} of $\Tprime$.}
\label{tab:TPrimeFSI}
\end{table}

As in our $\Delta(27)$ example, one could also construct an explicit example
model based on this group and calculate CP asymmetries. However, as has been
pointed out in \Secref{sec:TypeIICPtrafo} already, there  will be no CP
violation originating from the intrinsic properties of \Tprime, i.e.\ from the
Clebsch--Gordan coefficients. That is, unlike in the $\Delta(27)$ case, there is
the \CPgen\ transformation \eqref{eq:TprimeAutomorphism} available that ensures 
physical CP conservation. Of course, CP could be violated explicitly if there
were not enough field rephasing degrees of freedom to absorb all complex
coupling parameter phases. This is, however, not related to the group structure
of the model and, thus, is not discussed here. Instead, let us use \Tprime\ as
an example to discuss realizations of the CP transformation
\eqref{eq:TprimeAutomorphism} in different bases and comment on complications
which may arise in some of these bases.

For the matrices \UU[_i]{\rep{r}}, with which one has to multiply the
representation vectors in addition to the conjugation, \HLS\ 
obtain\footnote{Note that \HLS\ use the \Tprime\ basis of Feruglio et
al.~\cite[Appendix A]{Feruglio:2007uu}~(see also \Appref{app:IshimoriBasis}).}
\begin{subequations}\label{eq:HolthausenCP}
\begin{align}
 \rep[_i]{1} & ~\xmapsto[\mathrm{HLS}]{\CPgen}~ \omega^{i}\,\rep[_i]{1}^*\;,\quad (0\le i\le 2)
 \;,\label{eq:HolthausenCPa}\\
 \rep[_i]{2} & ~\xmapsto[\mathrm{HLS}]{\CPgen}~\diag(\psi^{-5},\psi^5)\,\rep[_i]{2}^*\;,\quad (0\le i\le 2)
 \;,\label{eq:HolthausenCPb}\\
 \rep{3} & ~\xmapsto[\mathrm{HLS}]{\CPgen}~ \diag(1,\omega,\omega^2)\,\rep{3}^* 
\end{align}
\end{subequations} 
with $\omega=\mathrm{e}^{2\pi\I/3}$, as before, and
$\psi=\mathrm{e}^{2\pi\I/24}$. This CP transformation is only unique up to
multiplication with \Tprime\ elements and can still be amended by phase factors
$\eta_{\ChargeC\ParityP}$ as in \Eqref{eq:CanonicalCP4fields} for each field.
However, this only corresponds to the freedom of rephasing fields, and,
therefore, one can choose a common phase for all fields in the same \Tprime\
representation without loss of generality. Yet, the specific choice made by
\HLS\ in combination with the Clebsch--Gordan coefficients of \cite[Appendix
A]{Feruglio:2007uu} appears inconvenient to us for the following reason.

Consider the contraction of $\psi$ in the representation \rep[_0]{2} with $\chi$
in the \rep[_1]{2} to the non--trivial singlet \rep[_1]{1},
\begin{equation}
 \left(\psi\otimes\chi\right)_{\rep[_1]{1}}
 ~=~\frac{-1}{\sqrt{2}}\,\left(\psi_1\,\chi_2-\psi_2\,\chi_1\right)\;.
\end{equation}
It is obvious that this contraction does not acquire a phase under the CP
transformation \eqref{eq:HolthausenCPb} because
\begin{equation}\label{eq:HLSdoublettrafo}
 \psi_1\,\chi_2-\psi_2\,\chi_1
 ~\xrightarrow{(\mathrm{\ref{eq:HolthausenCPb}})}~
 \psi_1^*\,\chi_2^*-\psi_2^*\,\chi_1^*\;. 
\end{equation}
However, according to \eqref{eq:HolthausenCPa}, the non--trivial singlet
\rep[_1]{1} should acquire a phase factor $\omega$. Hence, a composite state in
the \Tprime\ representation \rep[_1]{1} transforms differently under CP from an
elementary \rep[_1]{1} which, although not inconsistent, is certainly
inconvenient. Moreover, this complication is unnecessary as we have shown in
\Secref{sec:TypeIICPtrafo} because for type~II~A groups it is always
possible to fix the CP transformation phases and the phases of the
Clebsch--Gordan coefficients in such a way that the CP transformation behavior
of fields, composite and elementary alike, directly follows from their
transformation behavior under the discrete flavor group.

In the basis of Feruglio et al.\ with the Clebsch--Gordan phases chosen as shown
in \Appref{app:Tprime} this more convenient CP transformation takes the form
\begin{subequations}
\label{eq:CPFeruglio}
\begin{align}
  \rep[_i]{1} & ~\xmapsto{\CPgen}~ \rep[_i]{1}^*\;,\\
  \rep[_i]{2} & ~\xmapsto{\CPgen}~ \diag{(1, \mathrm{e}^{5\pi\,\I/6})} \, \rep[_i]{2}^*\;,
  \label{eq:Transformation2FBApp}\\
  \rep{3} & ~\xmapsto{\CPgen}~ \diag{(1,\omega,\omega^2)} \, \rep{3}^*\;.
\end{align}
\end{subequations}

Furthermore, it turns out that the chosen basis itself has a certain deficit.
The \rep{3} of \Tprime\ is a real representation. However, the corresponding
representation matrices in the Feruglio et al.\ basis are complex matrices which
means that one cannot consistently describe a field $\myvec\phi$ in the
three--dimensional representation by a triplet of manifestly real scalar
fields. One may ``rectify'' this by imposing a Majorana--like condition which constrains the complex
entries of the field to the correct number of degrees of freedom, i.e.\ 
\begin{equation}\label{eq:MajoranaLikeCondition}
  \myvec\phi^* ~=~ U\,\myvec\phi ~=~ \begin{pmatrix} 1 & 0 & 0\\ 0 & 0 & 1\\ 0 &
1 & 0 \end{pmatrix} \, \myvec\phi\;.
\end{equation}
This is also reflected by the kinetic term, which is given by
\begin{equation}\label{eq:nonpositivekineticTerm}
  \frac{1}{2}\,\left(\partial_\mu\myvec\phi \otimes \partial^\mu\myvec\phi\right)_{\rep[_0]{1}} 
  ~=~ 
  \frac{1}{2}\,\partial_\mu\myvec\phi^T \, U^T \, \partial^\mu\myvec\phi
\end{equation}
and is non--positive definite for real field values. If one instead imposes the
Majorana--like condition on the kinetic term,
\begin{equation}
  \frac{1}{2}\,\partial_\mu\myvec\phi^T \, U^T \, \partial^\mu\myvec\phi \stackrel{\eqref{eq:MajoranaLikeCondition}}{~=~} \frac{1}{2}\,\partial_\mu\myvec\phi^\dagger \, \partial^\mu\myvec\phi\;,
\end{equation}
one can see that for the now complex field values the kinetic term is indeed
positive definite.
Nonetheless, treating $\myvec\phi$ as a complex field and enforcing a condition
like \Eqref{eq:MajoranaLikeCondition} complicates perturbative computations,
and, for performing those, it appears more convenient to avoid such a situation
altogether by going to a basis where the triplet representation matrices are all
real.\footnote{If only complex fields are present in a model, like in
supersymmetric models, this issue does not arise.} Such a basis is the
``Ma--Rajasekaran basis'' \cite{Ma:2001dn} (see also \Appref{app:MaBasis}).
In this basis, the CP transformation is given by
\begin{subequations}\label{eq:CPtrafoTprimeMaApp}
\begin{align}
  \rep[_i]{1} & ~\xmapsto{\CPgen}~ \rep[_i]{1}^*\;,\\
  \rep[_i]{2} & ~\xmapsto{\CPgen}~ 
  \rep[_i]{2}^*\;,\\
  \rep{3} & ~\xmapsto{\CPgen}~ \begin{pmatrix}1 & 0 & 0 \\ 0 & 0 & 1 \\ 0 & 1 & 0\end{pmatrix} \, \rep{3}^*\;.
\end{align}
\end{subequations}

Since \Tprime\ is of type~II~A, the group also admits a basis with completely
real Clebsch--Gordan coefficients, which is given by the one of Ishimori et
al.~\cite{Ishimori:2010au} (see also \Appref{app:IshimoriBasis}) for the choice
$p=\I$ and $p_1=p_2=1$. In this basis, the generalized CP transformation is
identical to conjugation since the matrices \UU[_i]{\rep{r}} are all unit
matrices. However, the Ishimori basis again suffers from the issues with real
triplet fields. Whether one choice of basis or the other is more convenient
depends on the specific model at hand.

\subsection{Example for a type~II~B group: \texorpdfstring{$\boldsymbol{\Sigma(72)}$}{Sigma(72)}}
\label{sec:twoB}

The non--Abelian group $\Sigma(72)$ is an example for groups of type~II~B.
Information on the generators, characters and tensor product contractions of
this group can be found in \Appref{app:Sigma72}.

As one can check explicitly with the twisted Frobenius--Schur indicator,
$\Sigma(72)$ has no Bickerstaff--Damhus automorphism, and, therefore, there is
no basis of the group in which all Clebsch--Gordan coefficients are real.
However, the group is ambivalent, i.e.\ each conjugacy class contains with an
element $g$ also its inverse element $g^{-1}$, which makes any class--preserving
automorphism at the same time class--inverting. Thus, the identity map, which is
trivially an involution, can be used to define a consistent and
model--independent CP transformation.  The corresponding twisted
Frobenius--Schur indicators are shown in \Tabref{tab:Sigma72FSI}, where the $-1$
for the two--dimensional representation signals that this representation
transforms with an anti--symmetric matrix under the automorphism, which
implies that the transformation is not a BDA.
\begin{table}[t]
\centering
\begin{tabular}{|c|*{6}{c}|}
\hline
 $\rep R$ & \rep[_0]{1} & \rep[_1]{1} & \rep[_2]{1} & \rep[_3]{1} & \rep{2} & \rep{8} \\
\hline
$\TFS{\mathrm{id}}(\rep R)$ & $1$ & $1$ & $1$ & $1$ & $-1$ & $1$ \\
\hline
\end{tabular}
\caption{Twisted Frobenius--Schur indicators for the identity automorphisms of
$\Sigma(72)$.}
\label{tab:Sigma72FSI}
\end{table}
In the basis specified in \Appref{app:Sigma72}, the corresponding CP
transformation takes a very simple form. In fact, the identity automorphism
leads to a CP transformation that acts as
\begin{align}
 (M,N,P)~\xmapsto{~\text{id}~}~(M,N,P)       & 
 &\quad\curvearrowright\quad 
 \rep[_i]{1}~\xmapsto{\CPgen}~\rep[_i]{1}^*\;,
 \quad
 \rep{2}~\xmapsto{\CPgen}~\UU{\rep{2}}\,\rep{2}^*\;,
 \quad
 \rep{8}~\xmapsto{\CPgen}~\rep{8}^*
\end{align}
on the irreducible representations, where one should bear in mind that all
representations of ambivalent groups are (pseudo--)real. Hence, the CP
transformation acts as conjugation on all representations except the \rep{2},
which has to be conjugated and multiplied with the anti--symmetric matrix
\begin{equation}
  \UU{{\rep{2}}} ~=~ \begin{pmatrix}0 & 1\\ -1 & 0 \end{pmatrix}\;.
\end{equation}
Therefore, as described in sections~\ref{sec:ProperCP} and
\ref{sec:TypeIICPtrafo}, imposing this CP transformation as a symmetry enlarges
the flavor group by an additional \Z{2} factor to $\Sigma(72) \times \Z{2}$. The
additional symmetry generator acts trivially on all representations except for
the \rep{2}, on which it acts as
$V_{\rep{2}}=\UU{{\rep{2}}}\,\UU{{\rep{2}}}^*=-\mathbbm{1}$. Hence, this
additional \Z{2} forbids all terms which contain an odd number of fields in the
two--dimensional representation \rep{2}. These are terms like
\begin{equation}
  \mathscr{L} ~\supset~  c\,\left(\rep{2} \otimes \left(\rep{8} \otimes \rep{8}\right)_{\rep{2}} \right)_{\rep[_0]{1}}
\end{equation}
which are exactly the ones that cannot be made CP invariant by any choice of coupling $c$ or by the
addition of any other term. On the other hand, if all terms which are prohibited
by the \Z2 are absent, the discussion of CP violation works in complete analogy
to type~II~A groups.

\section{Conclusions}
\label{sec:Conclusions}

In this study, we have discussed CP transformations in settings with a
discrete (flavor) symmetry \DiscreteGroup. We have shown that physical CP
transformations are given by the class--inverting automorphisms of
\DiscreteGroup, which implies that canonical CP transformations necessarily have
to be generalized due to the nature of discrete groups. This can only be avoided
for certain groups (type II A) in very specific bases, or in models with a
non--generic field content.

One of the central results of our discussion is that there are discrete groups
that automatically violate CP in the sense that they do not allow us to impose a
consistent (generalized) CP transformation for a generic field content.

More specifically, we have shown that there are three types of discrete
groups:
\begin{description}
 \item[Type I:] Groups that, in general, violate CP. Such groups do not possess
 a class--inverting automorphism which would be necessary in order to define
 a (generalized) CP transformation that can warrant physical CP conservation. 
 In generic settings based on such groups CP is violated explicitly. 
 This statement does not apply to non--generic models. For instance,
 if a model contains only a subset of irreducible
 representations for which 
 an automorphism $u$ exists that exchanges each of these representations by its
 conjugate, one can impose the generalized CP transformation corresponding
 to $u$ to be a symmetry, and thus guarantee CP conservation.
 \item[Type II A:] Groups that do admit real Clebsch--Gordan coefficients. For
such groups one can always define a physical CP transformation and find a CP
basis.\footnote{We have also commented on the fact that, although bases with
real Clebsch--Gordan coefficients exist for type II~A groups, such bases
may not be the most convenient choices for performing computations. Possible
issues include that it may not be possible to represent real representations by
manifestly real fields. In such a case it might be advantageous to work in basis
with complex Clebsch--Gordan coefficients and to use a generalized CP
transformation with a non--trivial \UCP\ when imposing CP invariance.} Whether
or not CP is violated in settings with type II A symmetries depends on the
number of complex couplings  versus the number of free field redefinition
phases, and, hence, the situation is very similar to that of continuous
symmetry groups such as \SU{N}.
 \item[Type II B:] Groups that do not admit real Clebsch--Gordan coefficients
but possess a class--inverting automorphism that can be used to
define a generalized CP transformation. Apart from the obvious possibility that,
like in the case of a type~II~A symmetry group, CP can be violated explicitly
(or spontaneously), here CP violation can arise from the presence of operators
which are prohibited by an additional symmetry, which might be introduced when
imposing CP invariance. That is, these groups have the unusual property that CP
invariance requires certain couplings to vanish rather than just restricting the
phases of the coupling coefficients. However, unlike in the type~I case, CP can
be imposed regardless of the matter content of a model based on a type~II~B
group.
\end{description}

We have discussed how one can use the (extended) twisted Frobenius--Schur
indicator as a tool to categorize the automorphisms and, henceforth, the
discrete groups.

As we have seen in an explicit example, spontaneous breaking of CP with
calculable phases can be achieved in settings in which a type II group gets
broken spontaneously to a type I group. CP violation can then be attributed to
some complex Clebsch--Gordan coefficients. That is, the CP phases are
predicted by group theory.

Another central outcome of our analysis is that some of the transformations that
have been coined ``generalized CP transformations'' in the recent literature are
just outer automorphisms, which have, a priori, nothing to do with CP.  As we
have demonstrated, imposing such a ``generalized CP transformation'' does not
lead to physical CP conservation, but in many cases it enlarges the original
(flavor) symmetry \DiscreteGroup\ to a larger group such that the setting can no
longer be called a, say, $\Delta(27)$ model. That is, although proper
generalized CP transformations are outer automorphisms of \DiscreteGroup, in
general, outer automorphisms as such have nothing to do with CP invariance. For
instance, $\Delta(27)$ has several outer automorphisms but does not allow for a
consistent CP transformation in a generic setting. As we have discussed in
detail, this is because none of the automorphisms is class--inverting.

\subsection*{Acknowledgments}

We would like to thank Oleg Lebedev, Christian Staudt and Patrick Vaudrevange
for useful discussions. M.R.\  would like to thank the  UC Irvine, where part of
this work was done, for  hospitality. M.-C.C.\ would like to thank TU M\"unchen,
where part of the work  was done, for hospitality. This work was partially
supported by the DFG cluster  of excellence ``Origin and Structure of the
Universe'', the Graduiertenkolleg ``Particle Physics at the Energy Frontier of
New Phenomena'' by Deutsche Forschungsgemeinschaft (DFG) and the TUM Graduate
School. The work of M.-C.C.\ was supported, in part, by the U.S.\ National
Science  Foundation (NSF) under Grant No.\ PHY-0970173. The work of K.T.M.\ was 
supported, in part, by the U.S.\ Department of Energy (DoE) under Grant 
No. DEFG02-04ER41290. M.-C.C. and M.R.\ would like
to thank the Aspen Center for Physics for  hospitality and support. M.-.C.C.\
thanks the Galileo Galilei Institute for Theoretical Physics for the
hospitality.  This research was done in the context of the ERC  Advanced Grant
project ``FLAVOUR''~(267104).

\appendix

\section{Group theory}

In this appendix we collect information on the groups used in the main text:
\Tprime, $\Delta(27)$ and $\Sigma(72)$. Some of the details were obtained with the
help of GAP~\cite{GAP4} and the \textsc{Mathematica}--package
\textsc{Discrete}~\cite{Holthausen:2011vd}.

\subsection{Group theory of \texorpdfstring{$\boldsymbol{\Tprime}$}{T'}} 
\label{app:Tprime}

We start by discussing some basic facts on \Tprime\ and compare different
conventions used in the literature. 

\subsubsection{\texorpdfstring{$\boldsymbol{\Tprime}$}{T'} generators}

\Tprime\ is generated by the operations $S$ and $T$ where
\begin{equation}
 S^4~=~T^3~=~(S\,T)^3~=~e\;.
\end{equation}
There are seven irreducible representations,
$\rep[_i]{1}$, $\rep[_i]{2}$ and $\rep{3}$, where $0\le i \le 2$ and the
representations \rep[_1]{1} and \rep[_1]{2} are conjugate to \rep[_2]{1} and \rep[_2]{2}, respectively.

\begin{table}[!h!]
\centering
\begin{tabular}{|r|ccc|ccc|c|}
\hline
  & $\rep[_0]{1}$ & $\rep[_1]{1}$ & $\rep[_2]{1}$
 & $\rep[_0]{2}$ & $\rep[_1]{2}$ & $\rep[_2]{2}$
 & $\rep{3}$\\
\hline
 $S$ & $1$ & $1$ & $1$ & $S_{\rep[_0]{2}}$ & $S_{\rep[_1]{2}}$ & $S_{\rep[_2]{2}}$ 
 & $S_{\rep{3}}$ \\
 $T$ & 1 & $\omega$ & $\omega^2$ 
 & $T_{\rep[_0]{2}}$ & $T_{\rep[_1]{2}}$ & $T_{\rep[_2]{2}}$ 
 & $T_{\rep{3}}$ \\
 \hline
\end{tabular}
\caption{Representations of the \Tprime\ generators.}
\label{tab:TprimeGenerators}
\end{table}

\subsubsection{\texorpdfstring{$\boldsymbol{\Tprime}$}{T'} tensor products}

The \Tprime\ tensor product rules are
\begin{subequations}
\begin{align}
 \rep[_i]{2}\otimes\rep[_j]{2} & ~=~ \rep{3}\oplus
 \rep{1}_{i+j~\mathrm{mod}~3}\;,\\
 \rep[_i]{2}\otimes\rep{3} & ~=~ \rep[_0]{2}\oplus\rep[_1]{2}\oplus\rep[_2]{2}\;,\\
 \rep{3}\otimes\rep{3} & ~=~ \rep{3}_\mathrm{s}\oplus\rep{3}_\mathrm{a}\oplus\rep[_0]{1}\oplus\rep[_1]{1}\oplus\rep[_2]{1}\;.
\end{align}
\end{subequations}

\subsubsection{Ma--Rajasekaran basis for \texorpdfstring{$\boldsymbol{\Tprime}$}{T'}}
\label{app:MaBasis}

A basis in which the representation matrices for the real triplet representation
are manifestly real has been given by Ma and Rajasekaran \cite{Ma:2001dn} in the
case of \Afour. For \Tprime\ it is given by
\begin{subequations}
\allowdisplaybreaks
\begin{align}
 S_{\rep[_i]{2}}^\mathrm{M} & ~=~ -\frac{1}{\sqrt{3}} 
 \begin{pmatrix}
  \I & \sqrt{2}\,\I \\ \sqrt{2}\,\I & -\I 
 \end{pmatrix}
 \;,\qquad (i=0,1,2)\;,\\
 T_{\rep[_0]{2}}^\mathrm{M} & ~=~
 \begin{pmatrix}
  \omega^2 & 0 \\ 0 & \omega
 \end{pmatrix}\;,
 \qquad
 T_{\rep[_1]{2}}^\mathrm{M} ~=~
 \begin{pmatrix}
  1 & 0 \\ 0 & \omega^2
 \end{pmatrix}\;,
 \qquad
 T_{\rep[_2]{2}}^\mathrm{M}~=~
 \begin{pmatrix}
  \omega & 0 \\ 0 & 1
 \end{pmatrix}\;,\\
 S_{\rep{3}}^\mathrm{M} & ~=~
 \begin{pmatrix}
  1 & 0 & 0 \\
  0 & -1 & 0 \\
  0 & 0 & -1
 \end{pmatrix}\;,\quad
 T_{\rep{3}}^\mathrm{M} ~=~
 \begin{pmatrix}
  0 & 1 & 0 \\ 0 & 0 & 1 \\ 1 & 0 & 0
 \end{pmatrix}\;.
\end{align}
\label{eq:Ma_basis}
\end{subequations}
The tensor products with the correct phases and correct normalization read
\begin{subequations}
\begin{align}
  \left(x_{\rep{3}}\otimes y_{\rep{3}}\right)_{\rep[_0]{1}} & ~=~  \frac{x_{1}\,y_{1}+x_{2}\,y_{2}+x_{3}\,y_{3}}{\sqrt{3}}\;,\\
  \left(x_{\rep{3}}\otimes y_{\rep{3}}\right)_{\rep[_1]{1}} & ~=~  \frac{x_{1}\,y_{1}+\omega^2\,x_{2}\,y_{2}+\omega\,x_{3}\,y_{3}}{\sqrt{3}}\;,\\
  \left(x_{\rep{3}}\otimes y_{\rep{3}}\right)_{\rep[_2]{1}} & ~=~  \frac{x_{1}\,y_{1}+\omega\,x_{2}\,y_{2}+\omega^2\,x_{3}\,y_{3}}{\sqrt{3}}\;,\\
  \left(x_{\rep{3}}\otimes y_{\rep{3}}\right)_{\rep[_\mathrm{s}]{3}} & ~=~ \frac{1}{\sqrt{2}}\,
  \begin{pmatrix}
    x_{2}\, y_{3}+x_{3}\, y_{2}\\
    x_{1}\, y_{3}+x_{3}\, y_{1}\\
    x_{1}\, y_{2}+x_{2}\, y_{1}
  \end{pmatrix}
  \;,\\
  \left(x_{\rep{3}}\otimes y_{\rep{3}}\right)_{\rep[_\mathrm{a}]{3}} & ~=~ \frac{\I}{\sqrt{2}}\,
  \begin{pmatrix}
    x_{2}\, y_{3}-x_{3}\, y_{2}\\
    x_{3}\, y_{1}-x_{1}\, y_{3}\\
    x_{1}\, y_{2}-x_{2}\, y_{1}
  \end{pmatrix}
  \;,\\
  \left(\psi_{\rep[_i]{2}}\otimes\chi_{\rep[_j]{2}}\right)_{\rep[_{i+j}]{1}}
  & ~=~  \frac{-1}{\sqrt{2}}\,\left(\psi_1\,\chi_2-\psi_2\,\chi_1\right)\;, \\
  \left(\psi_{\rep[_i]{2}}\otimes\chi_{\rep[_{3-i}]{2}}\right)_{\rep{3}}
  & ~=~ 
  \frac{1}{\sqrt{3}}\,
  \begin{pmatrix}
          -\psi_1\chi_1+\frac{1}{\sqrt{2}}\,\left( \psi_1\, \chi_2 +\psi_2\, \chi_1\right)+ \psi_2\, \chi_2\\
          -\omega\,\psi_1\chi_1+\frac{1}{\sqrt{2}}\,\left( \psi_1\, \chi_2 + \psi_2\, \chi_1\right)+ \omega^2\,\psi_2\, \chi_2\\
          -\omega^2\,\psi_1\chi_1+\frac{1}{\sqrt{2}}\,\left( \psi_1\, \chi_2 + \psi_2\, \chi_1\right)+ \omega\,\psi_2\, \chi_2
  \end{pmatrix}
  \;,\\
  \left(\psi_{\rep[_i]{2}}\otimes\chi_{\rep[_{2-i}]{2}}\right)_{\rep{3}}
  & ~=~ 
  \frac{1}{\sqrt{3}}\,
  \begin{pmatrix}
          -\psi_1\chi_1+\frac{1}{\sqrt{2}}\,\left( \psi_1\, \chi_2
          +\psi_2\, \chi_1\right)+ \psi_2\, \chi_2\\
          -\psi_1\chi_1+\omega^2\,\frac{1}{\sqrt{2}}\,\left( \psi_1\, \chi_2
          +\psi_2\, \chi_1\right)+ \omega\,\psi_2\, \chi_2\\
          -\psi_1\chi_1+\omega\,\frac{1}{\sqrt{2}}\,\left( \psi_1\, \chi_2
          +\psi_2\, \chi_1\right)+ \omega^2\,\psi_2\, \chi_2
  \end{pmatrix}
  \;,\\
  \left(\psi_{\rep[_i]{2}}\otimes\chi_{\rep[_{1-i}]{2}}\right)_{\rep{3}}
  & ~=~
  \frac{1}{\sqrt{3}}\,
  \begin{pmatrix}
          -\psi_1\chi_1+\frac{1}{\sqrt{2}}\,\left( \psi_1\, \chi_2 + \psi_2\, \chi_1\right)+ \psi_2\, \chi_2\\
          -\omega^2\,\psi_1\chi_1+\omega\,\frac{1}{\sqrt{2}}\,\left( \psi_1\, \chi_2 + \psi_2\, \chi_1\right)+ \psi_2\, \chi_2\\
          -\omega\,\psi_1\chi_1+\omega^2\,\frac{1}{\sqrt{2}}\,\left( \psi_1\, \chi_2 + \psi_2\, \chi_1\right)+ \psi_2\, \chi_2
  \end{pmatrix}
  \;,\\
  \left(\psi_{\rep[_i]{2}}\otimes x_{\rep{3}}\right)_{\rep[_i]{2}}
  & ~=~ 
  \frac{1}{3}\,
  \begin{pmatrix}
          \psi_1\left(x_1+x_2+x_3\right)+\sqrt{2}\,\psi_2\left(x_1+\omega^2\,x_2+\omega\,x_3\right) \\
          \sqrt{2}\,\psi_1\left(x_1+\omega\,x_2+\omega^2\,x_3\right)-\psi_2\left(x_1+x_2+x_3\right)
  \end{pmatrix}
  \;,\\
  \left(\psi_{\rep[_i]{2}}\otimes x_{\rep{3}}\right)_{\rep[_{i+1}]{2}}
  & ~=~
  \frac{1}{3}\,
  \begin{pmatrix}
          \psi_1\left(x_1+\omega^2\,x_2+\omega\,x_3\right)+\sqrt{2}\,\psi_2\left(x_1+\omega\,x_2+\omega^2\,x_3\right) \\
          \sqrt{2}\,\psi_1\left(x_1+x_2+x_3\right)-\psi_2\left(x_1+\omega^2\,x_2+\omega\,x_3\right)
  \end{pmatrix}
  \;,\\
  \left(\psi_{\rep[_i]{2}}\otimes x_{\rep{3}}\right)_{\rep[_{i+2}]{2}}
  & ~=~
  \frac{1}{3}\,
  \begin{pmatrix}
          \psi_1\left(x_1+\omega\,x_2+\omega^2\,x_3\right)+\sqrt{2}\,\psi_2\left(x_1+x_2+x_3\right) \\
          \sqrt{2}\,\psi_1\left(x_1+\omega^2\,x_2+\omega\,x_3\right)-\psi_2\left(x_1+\omega\,x_2+\omega^2\,x_3\right)
  \end{pmatrix}
  \;.
\end{align}
\end{subequations}

\subsubsection{Ishimori et al.\ basis}
\label{app:IshimoriBasis}

Another basis for the triplet representation has been discussed in
\cite{Ishimori:2010au}, where one uses
\begin{subequations}
\allowdisplaybreaks
\label{eq:Ishimori_basis}
\begin{align}
 S_{\rep[_i]{2}}^\mathrm{I} & ~=~ -\frac{1}{\sqrt{3}} 
 \begin{pmatrix}
  \I & \sqrt{2}\,p \\ -\sqrt{2}\,p^* & -\I 
 \end{pmatrix}
 \;,\qquad (i=0,1,2)\;,\\
 T_{\rep[_0]{2}}^\mathrm{I} & ~=~
 \begin{pmatrix}
  \omega^2 & 0 \\ 0 & \omega
 \end{pmatrix}\;,
 \qquad
 T_{\rep[_1]{2}}^\mathrm{I} ~=~
 \begin{pmatrix}
  1 & 0 \\ 0 & \omega^2
 \end{pmatrix}\;,
 \qquad
 T_{\rep[_2]{2}}^\mathrm{I}~=~
 \begin{pmatrix}
  \omega & 0 \\ 0 & 1
 \end{pmatrix}\;,\\
 S_{\rep{3}}^\mathrm{I} & ~=~
 \begin{pmatrix}
  -1 & 2\, p_1 & 2\, p_1\, p_2 \\ 2\, p_1^* & -1 & 2 p_2 \\ 2\, p_1^*\, p_2^* & 2\, p_2^* & -1
 \end{pmatrix}
 \;,\\
 T_{\rep{3}}^\mathrm{I} & ~=~
 \begin{pmatrix}
  1 & 0 & 0 \\ 0 & \omega  & 0 \\ 0 & 0 & \omega ^2
 \end{pmatrix}
\end{align}
\end{subequations}
as generators with $p=\mathrm{e}^{\I\,\varphi}$,
$p_1=\mathrm{e}^{\I\,\varphi_1}$, and $p_2=\mathrm{e}^{\I\,\varphi_2}$, where
$\varphi$, $\varphi_1$, and $\varphi_2$ are arbitrary real phases.  The
free phases of the triplet representation can be removed by a transformation
$\widetilde{S}_{\rep{3}}^\mathrm{I}=P\,{S}_{\rep{3}}^\mathrm{I}\,P^\dagger$
with 
\begin{equation}
 P~=~
 \left(\begin{array}{ccc}
  1 & 0 & 0 \\
  0 & \mathrm{e}^{\I\,\varphi_1} & 0 \\
  0 & 0 & \mathrm{e}^{\I\,(\varphi_1+\varphi_2)}
 \end{array}\right)\;.
\end{equation}

The transformation which connects the bases \eqref{eq:Ma_basis} and
\eqref{eq:Ishimori_basis} for the triplet representations is given by
\begin{equation}\label{eq:basistrafo}
 S_{\rep{3}}^\mathrm{M}
 ~=~(\widetilde{U}\,P)\, S_{\rep{3}}^\mathrm{I}\,(\widetilde{U}\,P)^\dagger
 \quad\text{and}\quad
 T_{\rep{3}}^\mathrm{M}
 ~=~(\widetilde{U}\,P)\, T_{\rep{3}}^\mathrm{I}\,(\widetilde{U}\,P)^\dagger,
\end{equation}
with 
\begin{equation}\label{eq:Utilde}
 \widetilde{U}~=~\frac{1}{\sqrt{3}}
 \left(\begin{array}{ccc}
  1 & 1 & 1\\
  1 & \omega & \omega^2 \\
  1 & \omega^2 & \omega
 \end{array}\right)\;.
\end{equation}

Note that for the particular choice of $p=\I$ and $p_1=p_2=1$,  the
representation matrices of basis \eqref{eq:Ishimori_basis} fulfill the
Bickerstaff--Damhus equation \eqref{eq:BDAequation2} for the outer automorphism
$(S,T)\rightarrow(S^3,T^2)$. Hence, in this particular basis, all
Clebsch--Gordan coefficients are real. This has also been found in an explicit
computation \cite{Ishimori:2010au}.

Another basis commonly used in the literature is the one of  Feruglio et
al.~\cite[Appendix A]{Feruglio:2007uu}, which can be obtained from \eqref{eq:Ishimori_basis} by setting 
$p_1=p_2=\mathrm{e}^{2\pi\I/3}$ and $p=\mathrm{e}^{2\pi\I/24}$. We adjust the
global phases of the tensor product contractions in this basis such that
for the CP transformation \eqref{eq:CPFeruglio} compound states transform like
elementary states and obtain
\begin{subequations}
\begin{align}
  \left(x_{\rep{3}}\otimes y_{\rep{3}}\right)_{\rep[_0]{1}} & ~=~  \frac{x_1 \,y_1 + x_2 \,y_3 + x_3 \,y_2 }{\sqrt{3}}\;,\\
  \left(x_{\rep{3}}\otimes y_{\rep{3}}\right)_{\rep[_1]{1}} & ~=~  \frac{\omega \, (x_1 \,y_2 + x_2 \,y_1 + x_3 \,y_3)}{\sqrt{3}}\;,\\
  \left(x_{\rep{3}}\otimes y_{\rep{3}}\right)_{\rep[_2]{1}} & ~=~  \frac{\omega^2 \, (x_1 \,y_3 + x_2 \,y_2 + x_3 \,y_1)}{\sqrt{3}}\;,\\
  \left(x_{\rep{3}}\otimes y_{\rep{3}}\right)_{\rep[_\mathrm{s}]{3}} & ~=~ \frac{1}{\sqrt{6}}\,
  \begin{pmatrix}
    2 x_1 \,y_1 - x_3 \,y_2 - x_2 \,y_3\\
    - x_2 \,y_1 - x_1 \,y_2 + 2 x_3 \,y_3\\
    - x_3 \,y_1 + 2 x_2 \,y_2 - x_1 \,y_3
  \end{pmatrix}
  \;,\\
  \left(x_{\rep{3}}\otimes y_{\rep{3}}\right)_{\rep[_\mathrm{a}]{3}} & ~=~ \frac{1}{\sqrt{2}}\,
  \begin{pmatrix}
    x_{2}\, y_{3}-x_{3}\, y_{2}\\
    x_{1}\, y_{2}-x_{2}\, y_{1}\\
    x_{3}\, y_{1}-x_{1}\, y_{3}
  \end{pmatrix}
  \;,\\
  \left(\psi_{\rep[_i]{2}}\otimes\chi_{\rep[_j]{2}}\right)_{\rep[_{i+j}]{1}}
  & ~=~  \frac{\mathrm{e}^{7\I\,\pi/12}}{\sqrt{2}}\,\left(\psi_1\,\chi_2-\psi_2\,\chi_1\right)\;, \\
  \left(\psi_{\rep[_i]{2}}\otimes\chi_{\rep[_{3-i}]{2}}\right)_{\rep{3}}
  & ~=~ \I\,\omega^2\,
  \begin{pmatrix}
          \frac{1-\I}{2}\, \left(\psi_1 \,\chi_2+\psi_2 \,\chi_1\right)\\
          \I \, \psi_1 \,\chi_1\\
          \psi_2 \,\chi_2
  \end{pmatrix}
  \;,\\
  \left(\psi_{\rep[_i]{2}}\otimes\chi_{\rep[_{2-i}]{2}}\right)_{\rep{3}}
  & ~=~ \I\,
  \begin{pmatrix}
          \I\,\psi_1 \,\chi_1\\
          \psi_2 \,\chi_2\\
          \frac{1-\I}{2}\, \left(\psi_1 \,\chi_2+\psi_2 \,\chi_1\right)
  \end{pmatrix}
  \;,\\
  \left(\psi_{\rep[_i]{2}}\otimes\chi_{\rep[_{1-i}]{2}}\right)_{\rep{3}}
  & ~=~ \I\,\omega\,
  \begin{pmatrix}
          \psi_2 \,\chi_2\\
          \frac{1-\I}{2}\, \left(\psi_1 \,\chi_2+\psi_2 \,\chi_1\right)\\
          \I\, \psi_1 \,\chi_1
  \end{pmatrix}
  \;,\\
  \left(\psi_{\rep[_i]{2}}\otimes x_{\rep{3}}\right)_{\rep[_i]{2}}
  & ~=~ 
  \frac{1}{\sqrt{3}}\,
  \begin{pmatrix}
          \psi_1 \,\chi_1+(1+\I)\, \psi_2 \,\chi_2\\
          (1-\I) \, \psi_1 \,\chi_3-\psi_2 \,\chi_1
  \end{pmatrix}
  \;,\\
  \left(\psi_{\rep[_i]{2}}\otimes x_{\rep{3}}\right)_{\rep[_{i+1}]{2}}
  & ~=~
  \frac{\omega}{\sqrt{3}}\,
  \begin{pmatrix}
          \left(\psi_1 \,\chi_2+(1+\I)\, \psi_2 \,\chi_3\right)\\
          (1-\I)\,\psi_1 \,\chi_1-\psi_2 \,\chi_2
  \end{pmatrix}
  \;,\\
  \left(\psi_{\rep[_i]{2}}\otimes x_{\rep{3}}\right)_{\rep[_{i+2}]{2}}
  & ~=~
  \frac{\omega^2}{\sqrt{3}}\,
  \begin{pmatrix}
          \psi_1 \,\chi_3+(1+\I) \, \psi_2 \,\chi_1\\
          (1-\I)\,\psi_1 \,\chi_2 - \psi_2 \,\chi_3
  \end{pmatrix}
  \;.
\end{align}
\end{subequations}

\subsection{Group theory of \texorpdfstring{$\boldsymbol{\Delta(27)}$}{Delta(27)}}
\label{app:Delta27}

$\Delta(27)$ is generated by the operations $A$ and $B$, where
\begin{equation}
 A^3~=~B^3~=~\left(A\,B\right)^3~=~e\;.
\end{equation}
The conjugacy classes are given as
\begin{align}
C_{1a}& : \{e \}\;,               &  \nonumber\\
C_{3a}& : \{A, BAB^2, B^2AB \}\;,           & C_{3b} &: \{A^2, BA^2B^2, B^2A^2B \}\;,&\nonumber\\
C_{3c}& : \{B, ABA^2, A^2BA \}\;,           & C_{3d} &: \{B^2, AB^2A^2, A^2B^2A \}\;,&\nonumber\\
C_{3e}& : \{ABA, A^2B, BA^2 \}\;,           & C_{3f} &: \{BAB, B^2A, AB^2\}\;,&\nonumber\\
C_{3g}& : \{AB, BA, A^2BA^2 \}\;,           & C_{3h} &: \{AB^2A, A^2B^2, B^2A^2 \}\;,&\nonumber\\
C_{3i}& : \{AB^2ABA \}\;,                   & C_{3j} &: \{BA^2BAB\}\;.&
\label{eq:ccs}
\end{align}
There are eleven inequivalent irreducible representations $\rep[_i]{1}$,
$\rep{3}$, and $\rep{\bar{3}}$, where $0\le i \le 8$.  The character table is
given in \ref{tab:Delta27char}.
\begin{table}[t!]
\centering
\resizebox{\textwidth}{!}{\begin{tabular}{c|ccccccccccc|}
                      &  $C_{1a}$ & $C_{3a}$ & $C_{3b}$ & $C_{3c}$ & $C_{3d}$ & $C_{3e}$ & $C_{3f}$ & $C_{3g}$ & $C_{3h}$ & $C_{3i}$ & $C_{3j}$ \\
                      &  1 &  3 &  3 &  3 &  3 &  3 &  3 &  3 &   3 &  1 &   1 \\
$\Delta(27)$       & $e$ & $A$ & $A^2$ & $B$    & $B^2$      & $ABA$      & $BAB$      & $AB$       & $A^2B^2$   & $AB^2ABA$ & $BA^2BAB$ \\
\hline
 $\rep[_0]{1}$ & $1$ & $1$        & $1$        & $1$        & $1$        & $1$        & $1$        & $1$        & $1$        & $1$       & $1$ \\
 $\rep[_1]{1}$ & $1$ & $1$        & $1$        & $\omega^2$ & $\omega$   & $\omega^2$ & $\omega$   & $\omega^2$ & $\omega$   & $1$       & $1$ \\
 $\rep[_2]{1}$ & $1$ & $1$        & $1$        & $\omega$   & $\omega^2$ & $\omega$   & $\omega^2$ & $\omega$   & $\omega^2$ & $1$       & $1$ \\
 $\rep[_3]{1}$ & $1$ & $\omega^2$ & $\omega$   & $1$        & $1$        & $\omega$   & $\omega^2$ & $\omega^2$ & $\omega$   & $1$       & $1$ \\
 $\rep[_4]{1}$ & $1$ & $\omega^2$ & $\omega$   & $\omega^2$ & $\omega$   & $1$        & $1$        & $\omega$   & $\omega^2$ & $1$       & $1$ \\
 $\rep[_5]{1}$ & $1$ & $\omega^2$ & $\omega$   & $\omega$   & $\omega^2$ & $\omega^2$ & $\omega$   & $1$        & $1$        & $1$       & $1$ \\
 $\rep[_6]{1}$ & $1$ & $\omega$   & $\omega^2$ & $1$        & $1$        & $\omega^2$ & $\omega$   & $\omega$   & $\omega^2$ & $1$       & $1$ \\
 $\rep[_7]{1}$ & $1$ & $\omega$   & $\omega^2$ & $\omega^2$ & $\omega$   & $\omega$   & $\omega^2$ & $1$        & $1$        & $1$       & $1$ \\
 $\rep[_8]{1}$ & $1$ & $\omega$   & $\omega^2$ & $\omega$   & $\omega^2$ & $1$        & $1$        & $\omega^2$ & $\omega$   & $1$       & $1$ \\
 $\rep{3}$       & $3$ &  $0$   & $0$ & $0$ & $0$ & $0$ & $0$ & $0$ & $0$ & $3\omega^2$ & $3\omega$   \\
 $\rep{\bar{3}}$ & $3$ &  $0$   & $0$ & $0$ & $0$ & $0$ & $0$ & $0$ & $0$ & $3\omega$   & $3\omega^2$ \\
\hline
\end{tabular}}
\caption{Character table of $\Delta(27)$. We define $\omega:=\mathrm{e}^{2\pi\,\I/3}$. 
The conjugacy classes (c.c.) are labeled by the order of their elements and a
letter.  The second line gives the cardinality of the corresponding c.c.\ and
the third line gives a representative of the c.c.\ in the presentation specified
in the text.}
\label{tab:Delta27char}
\end{table}

We adopt the labeling of \cite{Holthausen:2012dk} with the difference that in
our notation $\rep[_i]{1}=\rep{1}^{\text{(HLS)}}_{i-1}$ and use the contractions
of  \cite{Ma:2006ip} translated to our conventions. There appears to be a
typographical error in the character table of \cite{Ma:2006ip} in which the
characters of the two conjugacy classes $C_{10}$ and $C_{11}$  should be
interchanged. The representations $(\rep1_1,\rep1_2)$, $(\rep1_3,\rep1_6)$,
$(\rep1_4,\rep1_8)$, and $(\rep1_5,\rep1_7)$ as well as the triplets are the
complex conjugate of each other. For the triplet $\rep{3}$ we use the 
representation matrices
\begin{equation}
 A~=~\begin{pmatrix} 0 & 1 & 0 \\ 0 & 0 & 1 \\ 1 & 0 & 0 \end{pmatrix}\;, \qquad B\,=\,\begin{pmatrix} 1 & 0 & 0 \\ 0 & \omega & 0 \\ 0 & 0 & \omega^2 \end{pmatrix}
\end{equation}
and for \rep{\bar{3}} the respective complex conjugate matrices.
This results in the multiplication rule
\begin{equation}
 x_{\rep{3}}\otimes \bar{y}_{\rep{\bar3}}~=~\sum^9_{i=1} \rep1_i\;,
\end{equation}
where
\begin{subequations}
\begin{align}
 \rep1_0 & ~=~\frac{\left(x_{1}\,\bar{y}_{1}+        x_{2}\,\bar{y}_{2}+         x_{3}\,\bar{y}_{3}\right)}{\sqrt{3}}\;,
 \\
 \rep1_1 & ~=~\frac{\left(x_{1}\,\bar{y}_{2}+        x_{2}\,\bar{y}_{3}+        
 x_{3}\,\bar{y}_{1}\right)}{\sqrt{3}}\;,&
 \rep1_2 & ~=~\frac{\left(x_{2}\,\bar{y}_{1}+        x_{3}\,\bar{y}_{2}+         x_{1}\,\bar{y}_{3}\right)}{\sqrt{3}}\;,
 \\
 \rep1_3 & ~=~\frac{\left(x_{1}\,\bar{y}_{1}+\omega\,  x_{2}\,\bar{y}_{2}
 +\omega^2\,x_{3}\,\bar{y}_{3}\right)}{\sqrt{3}}\;,&   
 \rep1_6 & ~=~\frac{\left(x_{1}\,\bar{y}_{1}+\omega^2\,x_{2}\,\bar{y}_{2}+\omega\,  
 x_{3}\,\bar{y}_{3}\right)}{\sqrt{3}}\;,
 \\
 \rep1_4 & ~=~\frac{\left(x_{1}\,\bar{y}_{2}+\omega\,  x_{2}\,\bar{y}_{3}
 +\omega^2\, x_{3}\,\bar{y}_{1}\right)}{\sqrt{3}}\;,&    
 \rep1_8 & ~=~\frac{\left(x_{2}\,\bar{y}_{1}+\omega^2\,x_{3}\,\bar{y}_{2}
 +\omega\,   x_{1}\,\bar{y}_{3}\right)}{\sqrt{3}}\;,
 \\  
 \rep1_5 & ~=~\frac{\left(x_{2}\,\bar{y}_{1}+\omega\,  x_{3}\,\bar{y}_{2}
 +\omega^2\, x_{1}\,\bar{y}_{3}\right)}{\sqrt{3}}\;,& 
 \rep1_7 & ~=~\frac{\left(x_{1}\,\bar{y}_{2}+\omega^2\,x_{2}\,\bar{y}_{3}
 +\omega\,   x_{3}\,\bar{y}_{1}\right)}{\sqrt{3}}\;. 
\end{align}
\end{subequations}
From the discussion in \Appref{app:ClassInverting} it is clear that $\Delta(27)$
as a non--Abelian group of odd order does not allow for any class--inverting
involutory automorphism.  Therefore, there is no possibility of having a
physical CP symmetry in a generic setup.  There are, however, several outer
automorphisms which exchange a subset of representations with their complex
conjugates.  If one is to construct a model with one of those subsets there is
the possibility of imposing physical CP conservation. Altogether there are 46
involutory automorphisms. Some examples are given by
\begin{subequations}
\begin{align}
 {u_1} & : (A,B)\,\rightarrow\,(A, B^2)           & &\curvearrowright\, \rep[_1]{1}\leftrightarrow\rep[_2]{1}\;,\;\rep[_4]{1}\leftrightarrow\rep[_5]{1}\;,\;\rep[_7]{1}\leftrightarrow\rep[_8]{1}\;,\;\rep3\rightarrow U_{{u_1}}\,\rep3^*\;, \\
 {u_2} & : (A,B)\,\rightarrow\,(ABA, B)           & &\curvearrowright\, \rep[_1]{1}\leftrightarrow\rep[_4]{1}\;,\;\rep[_2]{1}\leftrightarrow\rep[_8]{1}\;,\;\rep[_3]{1}\leftrightarrow\rep[_6]{1}\;,\;\rep3\rightarrow U_{{u_2}}\,\rep3^*\;, \\
 {u_3} & : (A,B)\,\rightarrow\,(BAB, B^2)         & &\curvearrowright\, \rep[_1]{1}\leftrightarrow\rep[_8]{1}\;,\;\rep[_2]{1}\leftrightarrow\rep[_4]{1}\;,\;\rep[_5]{1}\leftrightarrow\rep[_7]{1}\;,\;\rep3\rightarrow U_{{u_3}}\,\rep3^*\;, \\
 {u_4} & : (A,B)\,\rightarrow\,(AB^2A, B)         & &\curvearrowright\, \rep[_1]{1}\leftrightarrow\rep[_7]{1}\;,\;\rep[_2]{1}\leftrightarrow\rep[_5]{1}\;,\;\rep[_3]{1}\leftrightarrow\rep[_6]{1}\;,\;\rep3\rightarrow U_{{u_4}}\,\rep3^*\;, \\
 {u_5} & : (A,B)\,\rightarrow\,(BA^2B^2,AB^2A^2)  & &\curvearrowright\, \rep[_i]{1}\leftrightarrow\rep[_i]{1}^*\;,\;\rep3\rightarrow U_{{u_5}}\,\rep3\;.
 \label{eq:u_5}
\end{align}
\end{subequations}
All other representations stay inert under the transformation, and by
$\rep3\rightarrow U_{u_i}\,\rep3^*$ we mean that fields in the triplet
representation  have to be multiplied by the corresponding matrix in addition to
a possible conjugation. In our basis, these matrices are given by
\begin{align}
 U_{{u_1}} & ~=~
\mathbbm{1}\;,&
\quad
U_{{u_2}} & ~=~
\begin{pmatrix}
 \omega & 0 & 0 \\
 0 & 0 & 1  \\
 0 & 1 & 0 \\
\end{pmatrix}\;,&
U_{{u_3}} & ~=~
\begin{pmatrix}
 1 & 0 & 0 \\
 0 & \omega^2 & 0  \\
 0 & 0 & \omega^2 \\
\end{pmatrix}\;,& 
\nonumber\\
 U_{{u_4}} & ~=~
\begin{pmatrix}
 1 & 0 & 0 \\
 0 & 0 & \omega  \\
 0 & \omega & 0 \\
\end{pmatrix}\;,&
\quad
U_{{u_5}} & ~=~
\begin{pmatrix}
 0 & 0 & \omega^2 \\
 0 & 1 & 0  \\
 \omega & 0 & 0 \\
\end{pmatrix}\;.&
\label{eq:UDelta27}
\end{align}
The corresponding twisted Frobenius--Schur indicators for all representations are given in \Tabref{tab:Delta27FSI}.
\begin{table}[!h!]
\centering
\begin{tabular}{|c|ccccccccccc|}
\hline
 $\rep R$ & \rep[_0]{1} & \rep[_1]{1} & \rep[_2]{1} & \rep[_3]{1} & \rep[_4]{1}
 & \rep[_5]{1} & \rep[_6]{1} & \rep[_7]{1} & \rep[_8]{1} & \rep{3} &
 \rep{\bar{3}} \\
\hline
$\TFS{{u_1}}(\rep R)$ & 1 & 1 & 1 & 0 & 0 & 0 & 0 & 0 & 0 & 1 & 1 \\
$\TFS{{u_2}}(\rep R)$ & 1 & 0 & 0 & 1 & 0 & 0 & 1 & 0 & 0 & 1 & 1 \\
$\TFS{{u_3}}(\rep R)$ & 1 & 0 & 0 & 0 & 0 & 1 & 0 & 1 & 0 & 1 & 1 \\
$\TFS{{u_4}}(\rep R)$ & 1 & 0 & 0 & 1 & 0 & 0 & 1 & 0 & 0 & 1 & 1 \\
$\TFS{{u_5}}(\rep R)$ & 1 & 1 & 1 & 1 & 1 & 1 & 1 & 1 & 1 & 0 & 0 \\
\hline
\end{tabular}
\caption{Twisted Frobenius--Schur indicators for some outer automorphisms of $\Delta(27)$.}
\label{tab:Delta27FSI}
\end{table}
One can convince oneself by computing all \FSI's that for models with fields in
more than two non--trivial one--dimensional representations and a triplet it is
impossible to find an automorphism that leads to a consistent CP transformation.
This is a highly interesting feature of this group.

\subsection{Group theory of \texorpdfstring{$\boldsymbol{\Sigma(72)}$}{Sigma(72)}}
\label{app:Sigma72}

The non--Abelian group $\Sigma(72)$ is isomorphic to the semi--direct product
group $(Z_3 \times Z_3) \SemiDirect Q_8$, where $Q_8$ is the quaternion
group, and is generated by three generators $M$, $N$ and $P$, which fulfill
the relations
\begin{align}
  & M^4 ~=~ N^4 ~=~ P^3 ~=~ \left(M^2\,P^{-1}\right)^2 ~=~ e\;, \quad M^2 ~=~ N^2\;, \quad M^{-1}\,N ~=~ N\,M\;,\notag\\
  & P\,M\,P\,N^{-1}\,M\,P^{-1}\,N ~=~ e\;, \quad N\,P\,M^{-1}\,P ~=~M\,P\,N\;.
\end{align}

$\Sigma(72)$ has $6$ inequivalent irreducible representations: four
one--dimensional (\rep[_{0-3}]{1}), one two--dimensional (\rep{2}), and one
eight--dimensional (\rep{8}). The characters of $\Sigma(72)$ are shown in
\Tabref{tab:Sigma72char}.
\begin{table}[t]
\centering
\begin{tabular}{c|*{6}{c}|}
                &  $C_{1a}$ & $C_{3a}$ & $C_{2a}$ & $C_{4a}$ & $C_{4b}$ & $C_{4c}$ \\
                &  1 &  8 &  9 &  18 &  18 &  18 \\
  $\Sigma(72)$   & $e$ & $P$ & $M^2$ & $M\,N$ & $N$ & $M$ \\
  \hline
  $\rep[_0]{1}$ & $1$ & $1$  & $1$  & $1$  & $1$  & $1$ \\
  $\rep[_1]{1}$ & $1$ & $1$  & $1$  & $1$  & $-1$ & $-1$\\
  $\rep[_2]{1}$ & $1$ & $1$  & $1$  & $-1$ & $1$  & $-1$\\
  $\rep[_3]{1}$ & $1$ & $1$  & $1$  & $-1$ & $-1$ & $1$ \\
  $\rep{2}$     & $2$ & $2$  & $-2$ & $0$  & $0$  & $0$ \\
  $\rep{8}$     & $8$ & $-1$ & $0$  & $0$  & $0$  & $0$ \\
  \hline
\end{tabular}
\caption{Character table of $\Sigma(72)$. The conjugacy classes (c.c.) are
labeled by the order of their elements and a letter.  The second line gives the
cardinality of the corresponding c.c.\ and the third line gives a representative
of the c.c.\ in the presentation specified in the text.}
\label{tab:Sigma72char}
\end{table}
From the character table one can also read off the matrix realizations of the generators for the one--dimensional representations. The generators for the two--dimensional representation are given by
\begin{equation*}
  M_{\rep{2}} ~=~ \begin{pmatrix}0 & 1\\ -1 & 0 \end{pmatrix}\;, \quad
  N_{\rep{2}} ~=~ \begin{pmatrix}-\I & 0\\ 0 & \I \end{pmatrix}\;, \quad
  P_{\rep{2}} ~=~ \begin{pmatrix}1 & 0\\ 0 & 1 \end{pmatrix}\;,\notag\\
\end{equation*}
and the three matrices
\begin{align*}
  M_{\rep{8}} & ~=~ \begin{pmatrix} 
    0 & 0 & 1 & 0 & 0 & 0 & 0 & 0 \\
    0 & 0 & 0 & 1 & 0 & 0 & 0 & 0 \\
    1 & 0 & 0 & 0 & 0 & 0 & 0 & 0 \\
    0 & -1 & 0 & 0 & 0 & 0 & 0 & 0 \\
    0 & 0 & 0 & 0 & 0 & 0 & 1 & 0 \\
    0 & 0 & 0 & 0 & 0 & 0 & 0 & 1 \\
    0 & 0 & 0 & 0 & 1 & 0 & 0 & 0 \\
    0 & 0 & 0 & 0 & 0 & -1 & 0 & 0
  \end{pmatrix}\;, ~
  N_{\rep{8}} ~=~ \begin{pmatrix} 
    0 & 0 & 0 & 0 & 1 & 0 & 0 & 0 \\
    0 & 0 & 0 & 0 & 0 & 1 & 0 & 0 \\
    0 & 0 & 0 & 0 & 0 & 0 & 1 & 0 \\
    0 & 0 & 0 & 0 & 0 & 0 & 0 & -1 \\
    1 & 0 & 0 & 0 & 0 & 0 & 0 & 0 \\
    0 & -1 & 0 & 0 & 0 & 0 & 0 & 0 \\
    0 & 0 & 1 & 0 & 0 & 0 & 0 & 0 \\
    0 & 0 & 0 & 1 & 0 & 0 & 0 & 0
  \end{pmatrix}\;,\notag\\
  P_{\rep{8}} & ~=~ \frac{1}{2}\,\begin{pmatrix}
     -1 &  \sqrt{3} & 0 & 0 & 0 & 0 & 0 & 0 \\
    -\sqrt{3} &  -1 & 0 & 0 & 0 & 0 & 0 & 0 \\
    0 & 0 & 2 & 0 & 0 & 0 & 0 & 0 \\ 
    0 & 0 & 0 & 2 & 0 & 0 & 0 & 0 \\
    0 & 0 & 0 & 0 &  -1 & -\sqrt{3} & 0 & 0 \\
    0 & 0 & 0 & 0 &  \sqrt{3} &  -1 & 0 & 0 \\
    0 & 0 & 0 & 0 & 0 & 0 &  -1 &  -\sqrt{3} \\
    0 & 0 & 0 & 0 & 0 & 0 &  \sqrt{3} &  -1
  \end{pmatrix}\;,
\end{align*}
generate the eight--dimensional representation.

The (non--trivial) tensor product contractions of $\Sigma(72)$ are the following:
\begin{subequations}
\begin{align*}
  \left(x_{\rep[_0]{1}}\otimes y_{\rep{2}}\right)_{\rep{2}} & ~=~ \frac{1}{\sqrt{2}}\,\begin{pmatrix}x_1\, y_1 \\ x_1 \,y_2\end{pmatrix}\;,&
  \left(x_{\rep[_1]{1}}\otimes y_{\rep{2}}\right)_{\rep{2}} & ~=~ \frac{\I}{\sqrt{2}}\,\begin{pmatrix}x_1\, y_2\\ x_1\, y_1\end{pmatrix}\;,\\
  \left(x_{\rep[_2]{1}}\otimes y_{\rep{2}}\right)_{\rep{2}} & ~=~ \frac{\I}{\sqrt{2}}\,\begin{pmatrix}x_1\, y_1\\ -x_1\, y_2\end{pmatrix}\;,&
  \left(x_{\rep[_3]{1}}\otimes y_{\rep{2}}\right)_{\rep{2}} & ~=~ \frac{1}{\sqrt{2}}\,\begin{pmatrix}x_1\, y_2\\ -x_1\, y_1\end{pmatrix}\;,
\end{align*}
\begin{align*}
  \left(x_{\rep{2}}\otimes y_{\rep{2}}\right)_{\rep[_0]{1}} & ~=~ \frac{1}{\sqrt{2}}\left(x_1\, y_2 - x_2\, y_1 \right)\;,&
  \left(x_{\rep{2}}\otimes y_{\rep{2}}\right)_{\rep[_1]{1}} & ~=~ \frac{\I}{\sqrt{2}}\left(x_1\, y_1 - x_2\, y_2 \right)\;,\\
  \left(x_{\rep{2}}\otimes y_{\rep{2}}\right)_{\rep[_2]{1}} & ~=~ \frac{\I}{\sqrt{2}}\left(x_1\, y_2 + x_2\, y_1 \right)\;,&
  \left(x_{\rep{2}}\otimes y_{\rep{2}}\right)_{\rep[_3]{1}} & ~=~ \frac{1}{\sqrt{2}}\left(x_1\, y_1 + x_2\, y_2 \right)\;,
\end{align*}
\begin{align*}
  \left(x_{\rep[_0]{1}}\otimes y_{\rep{8}}\right)_{\rep{8}} & ~=~ \left(x_1\, y_1, \, x_1\, y_2, \, x_1\, y_3, \, x_1\, y_4, \, x_1\, y_5, \, x_1\, y_6, \, x_1\, y_7, \, x_1\, y_8\right)^T\;,\\
  \left(x_{\rep[_1]{1}}\otimes y_{\rep{8}}\right)_{\rep{8}} & ~=~ \left(x_1\, y_1, \, x_1\, y_2, \, -x_1\, y_3, \, -x_1\, y_4, \, -x_1\, y_5, \, -x_1\, y_6, \, x_1\, y_7, \, x_1\, y_8\right)^T\;,\\
  \left(x_{\rep[_2]{1}}\otimes y_{\rep{8}}\right)_{\rep{8}} & ~=~ \left(x_1\, y_1,\, x_1\, y_2,\, -x_1\, y_3,\, -x_1\, y_4,\, x_1\, y_5,\, x_1\, y_6,\, -x_1\, y_7,\, -x_1\, y_8\right)^T\;,\\
  \left(x_{\rep[_3]{1}}\otimes y_{\rep{8}}\right)_{\rep{8}} & ~=~ \left(x_1\, y_1,\, x_1\, y_2,\, x_1\, y_3,\, x_1\, y_4,\, -x_1\, y_5,\, -x_1\, y_6,\, -x_1\, y_7,\, -x_1\, y_8\right)^T\;,\\
  \left(x_{\rep{2}}\otimes y_{\rep{8}}\right)_{\rep{8}^1} & ~=~ \left(\I\, x_1\, y_2,\, -\I\, x_1\, y_1,\, \I\, x_2\, y_4,\, -\I\, x_2\, y_3,\, x_1\, y_6,\, -x_1\, y_5,\, x_2\, y_8,\, -x_2\, y_7\right)^T\;,\\
  \left(x_{\rep{2}}\otimes y_{\rep{8}}\right)_{\rep{8}^2} & ~=~ \left(\I\, x_2\, y_2,\, -\I\, x_2\, y_1,\, -\I\, x_1\, y_4,\, \I\, x_1\, y_3,\, -x_2\, y_6,\, x_2\, y_5,\, x_1\, y_8,\, -x_1\, y_7\right)^T\;,\\
  \left(x_{\rep{8}}\otimes y_{\rep{8}}\right)_{\rep[_0]{1}} & ~=~ \frac{1}{2\sqrt{2}}\,\left(x_1\, y_1 + x_2\, y_2 + x_3\, y_3 + x_4\, y_4 + x_5\, y_5 + x_6\, y_6 + x_7\, y_7 + x_8\, y_8\right)\;,\\
  \left(x_{\rep{8}}\otimes y_{\rep{8}}\right)_{\rep[_1]{1}} & ~=~ \frac{1}{2\sqrt{2}}\,\left(x_1\, y_1 + x_2\, y_2 - x_3\, y_3 - x_4\, y_4 - x_5\, y_5 - x_6\, y_6 + x_7\, y_7 + x_8\, y_8\right)\;,\\
  \left(x_{\rep{8}}\otimes y_{\rep{8}}\right)_{\rep[_2]{1}} & ~=~ \frac{1}{2\sqrt{2}}\,\left(x_1\, y_1 + x_2\, y_2 - x_3\, y_3 - x_4\, y_4 + x_5\, y_5 + x_6\, y_6 - x_7\, y_7 - x_8\, y_8\right)\;,\\
  \left(x_{\rep{8}}\otimes y_{\rep{8}}\right)_{\rep[_3]{1}} & ~=~ \frac{1}{2\sqrt{2}}\,\left(x_1\, y_1 + x_2\, y_2 + x_3\, y_3 + x_4\, y_4 - x_5\, y_5 - x_6\, y_6 - x_7\, y_7 - x_8\, y_8\right)\;,\\
  \left(x_{\rep{8}}\otimes y_{\rep{8}}\right)_{\rep{2}^1} & ~=~ \frac{1}{2}\,\begin{pmatrix}\I\, x_2\, y_1 - \I\, x_1\, y_2 - x_6\, y_5 + x_5\, y_6 \\ \I\, x_4\, y_3 - \I\, x_3\, y_4 - x_8\, y_7 + x_7\, y_8\end{pmatrix}\;,\\
  \left(x_{\rep{8}}\otimes y_{\rep{8}}\right)_{\rep{2}^2} & ~=~ \frac{1}{2}\,\begin{pmatrix}\I\, x_4\, y_3 - \I\, x_3\, y_4 + x_8\, y_7 - x_7\, y_8 \\ -\I\, x_2\, y_1 + \I\, x_1\, y_2 - x_6\, y_5 + x_5\, y_6\end{pmatrix}\;,\\
  \left(x_{\rep{8}}\otimes y_{\rep{8}}\right)_{\rep{8}^1} & ~=~ \frac{1}{\sqrt{2}}\,\left(x_1\, y_1 - x_2\, y_2,\,-x_2\, y_1 - x_1\, y_2,\,x_3\, y_3 - x_4\, y_4,\,-x_4\, y_3 - x_3\, y_4,\right.\notag\\
  & \qquad\qquad \left.x_5\, y_5 - x_6\, y_6,\,-x_6\, y_5 - x_5\, y_6,\,x_7\, y_7 - x_8\, y_8,\,-x_8\, y_7 - x_7\, y_8\right)^T\;,\\
  \left(x_{\rep{8}}\otimes y_{\rep{8}}\right)_{\rep{8}^2} & ~=~ \frac{1}{\sqrt{2}}\,\left(x_3\, y_5 + x_4\, y_6,\,x_4\, y_5 - x_3\, y_6,\,x_1\, y_7 - x_2\, y_8,\,-x_2\, y_7 - x_1\, y_8,\right.\notag\\
  & \qquad\qquad \left.x_7\, y_1 + x_8\, y_2,\,-x_8\, y_1 + x_7\, y_2,\,x_5\, y_3 - x_6\, y_4,\,x_6\, y_3 + x_5\, y_4\right)^T\;,\\
  \left(x_{\rep{8}}\otimes y_{\rep{8}}\right)_{\rep{8}^3} & ~=~ \frac{1}{\sqrt{2}}\,\left(x_3\, y_7 - x_4\, y_8,\,-x_4\, y_7 - x_3\, y_8,\,x_1\, y_5 - x_2\, y_6,\,x_2\, y_5 + x_1\, y_6,\right.\notag\\
  & \qquad\qquad \left.x_7\, y_3 + x_8\, y_4,\,x_8\, y_3 - x_7\, y_4,\,x_5\, y_1 + x_6\, y_2,\,-x_6\, y_1 + x_5\, y_2\right)^T\;,\\
  \left(x_{\rep{8}}\otimes y_{\rep{8}}\right)_{\rep{8}^4} & ~=~ \frac{1}{\sqrt{2}}\,\left(x_5\, y_7 - x_6\, y_8,\,x_6\, y_7 + x_5\, y_8,\,x_7\, y_5 + x_8\, y_6,\,x_8\, y_5 - x_7\, y_6,\right.\notag\\
  & \qquad\qquad \left.x_1\, y_3 + x_2\, y_4,\,-x_2\, y_3 + x_1\, y_4,\,x_3\, y_1 - x_4\, y_2,\,-x_4\, y_1 - x_3\, y_2\right)^T\;,\\
  \left(x_{\rep{8}}\otimes y_{\rep{8}}\right)_{\rep{8}^5} & ~=~ \frac{1}{\sqrt{2}}\,\left(x_5\, y_3 + x_6\, y_4,\,-x_6\, y_3 + x_5\, y_4,\,x_7\, y_1 - x_8\, y_2,\,-x_8\, y_1 - x_7\, y_2,\right.\notag\\
  & \qquad\qquad \left.x_1\, y_7 + x_2\, y_8,\,x_2\, y_7 - x_1\, y_8,\,x_3\, y_5 - x_4\, y_6,\,x_4\, y_5 + x_3\, y_6\right)^T\;,\\
  \left(x_{\rep{8}}\otimes y_{\rep{8}}\right)_{\rep{8}^6} & ~=~ \frac{1}{\sqrt{2}}\,\left(x_7\, y_5 - x_8\, y_6,\,x_8\, y_5 + x_7\, y_6,\,x_5\, y_7 + x_6\, y_8,\,-x_6\, y_7 + x_5\, y_8,\right.\notag\\
  & \qquad\qquad \left.x_3\, y_1 + x_4\, y_2,\,x_4\, y_1 - x_3\, y_2,\,x_1\, y_3 - x_2\, y_4,\,-x_2\, y_3 - x_1\, y_4\right)^T\;,\\
  \left(x_{\rep{8}}\otimes y_{\rep{8}}\right)_{\rep{8}^7} & ~=~ \frac{1}{\sqrt{2}}\,\left(x_7\, y_3 - x_8\, y_4,\,-x_8\, y_3 - x_7\, y_4,\,x_5\, y_1 - x_6\, y_2,\,x_6\, y_1 + x_5\, y_2,\right.\notag\\
  & \qquad\qquad \left.x_3\, y_7 + x_4\, y_8,\,-x_4\, y_7 + x_3\, y_8,\,x_1\, y_5 + x_2\, y_6,\,x_2\, y_5 - x_1\, y_6\right)^T
  \;.
\end{align*}
\end{subequations}

\section{A GAP code to compute the twisted Frobenius--Schur indicator}
\label{app:GAP}

The following code for GAP computes the twisted
Frobenius--Schur indicators for all irreducible representations of a finite
group \texttt{G} and a given automorphism \texttt{aut} of this group.
\begin{verbatim}
  twistedFS:=function(G,aut)
    local elG,tbl,irr,fsList;
    elG:=Elements(G);
    tbl:=CharacterTable(G);
    irr:=Irr(tbl);
    fsList:=List(elG,x->x*x^aut);
    return List(irr,y->Sum(fsList,x->x^y))/Size(G);
  end;
\end{verbatim}
This code can easily be augmented to compute the
$n^\mathrm{th}$ extended twisted Frobenius--Schur indicator (see
\Appref{app:extendedFS}). However, due to the large number of group operations,
the computation can be very time--consuming. In practice, it might,
therefore, be advisable to check directly whether a given high--order
automorphisms is class--inverting instead of using the $\FSI^{(n)}$.

As an example, one can print out the twisted Frobenius--Schur indicators of a
certain group for all its involutory automorphisms. In the example below,
the group $\text{SG}(24,3)$, i.e.\ \Tprime, is chosen.
\begin{verbatim}
  G:=SmallGroup(24,3);;
  autG:=AutomorphismGroup(G);;
  elAutG:=Elements(autG);;
  ordTwoAut:=Filtered(elAutG,x->Order(x)<=2);;
  for i in ordTwoAut do Print(twistedFS(G,i)); od;
\end{verbatim}

\section{Class--inverting automorphisms}
\label{app:ClassInverting}

An automorphism $u$ is class--inverting if and only if it sends each group
element to the conjugacy class of its inverse, i.e.\
\begin{equation}
  \forall~g\in \DiscreteGroup~:~\exists~h\in  \DiscreteGroup~:~u(g)~=~h \,g^{-1}\, h^{-1}\;.
\end{equation}
Class--inverting automorphisms have always even order except in ambivalent
groups, where the notions of class--inverting and class--preserving coincide and
where also odd--order automorphisms can be class--inverting.

In this appendix, we state some of the proofs omitted in the main text. The
discussion is in parts similar to \cite{Nishi:2013jqa}, where, however, the
assumptions were slightly different.

\subsection{Higher--order class--inverting automorphisms}
\label{app:higherOrder}

As stated in the main text, higher--order class--inverting automorphisms,
where higher--order means greater than two, do not seem to play any role as CP
transformations. Indeed, for all groups of order less than 150, with the
exception of order 128, we have checked that all such automorphisms are related
to order--two class--inverting automorphisms via inner automorphisms. That is,
let $u$ be a class--inverting automorphism which is of order greater than two,
i.e.\ which is not involutory. Moreover, assume that it squares to an inner
automorphism,
\begin{equation}\label{eq:usquare}
  \exists ~a \in \DiscreteGroup~:~u^2(g) 
  ~=~ 
  a \, g \, a^{-1} \quad \forall~ g \in \DiscreteGroup\;,
\end{equation}
for the reasons given in \Secref{sec:ProperCP}. Then, in all the checked
examples, there is a second automorphism $u'$ such that
\begin{subequations}
\begin{align}
  \label{eq:innerRelation}u'(g) & ~=~ b\, u(g)\, b^{-1} \quad 
  \text{with}~
  b \in \DiscreteGroup~\text{and}~\forall ~g \in \DiscreteGroup\;,\\
  u'^2(g) & ~=~ g \quad \forall ~g \in \DiscreteGroup\;.
\end{align}
\end{subequations}
Automorphisms $u$ and $u'$ which are related in this way lead to physically
equivalent CP transformations because
\begin{equation}
  \UU[_i]{\rep{r}}' ~=~ \rhoR{_i}(b)\,\UU[_i]{\rep{r}}
\end{equation}
for all irreducible representations \rep[_i]{r}, where \UU[_i]{\rep{r}} solves
\Eqref{eq:consistency1} for $u$ and $\UU[_i]{\rep{r}}'$ for $u'$,
respectively.\footnote{This implies that  $V_{\rep[_i]{r}}' =
\rhoR{_i}(b\,u(b))\,V_{\rep[_i]{r}}$. \label{fn:Vinner}} The condition that $u'$ is involutory,
\begin{align}
  \notag u'^2(g) & ~=~ b \,u\left(u'(g)\right) \, b^{-1} 
  ~=~ b \, u\left(b \, u(g) \, b^{-1}\right) \, b^{-1}\\
  & ~=~ b \,u(b) \,a\, g\, a^{-1}\, u(b)^{-1} \,b^{-1} 
  ~\stackrel{!}=~{g} \qquad\qquad\qquad \forall ~g \in \DiscreteGroup\;,
\end{align}
is fulfilled if and only if
\begin{equation}\label{eq:centrecondition}
  b \, u(b) \,  a ~\in~ Z(\DiscreteGroup)\;,
\end{equation}
where $Z(\DiscreteGroup)$ is the center of \DiscreteGroup. We will prove that,
for certain classes of automorphisms and groups, one can always find a group
element $b$ such that \eqref{eq:centrecondition} is fulfilled, i.e.\ that these
automorphisms are physically equivalent to involutory automorphisms.

As a first step, we show using representation theory that $u(a)=c\,a$ with an
appropriate $c\in Z(G)$, where $Z(G)$ is the center of \DiscreteGroup.  To see
this, equate the action of $u^2$ on the matrix realization of an irreducible
representation \rep[_i]{r} according to \Eqref{eq:consistency1} with
\Eqref{eq:usquare},
\begin{equation}
 \rhoR{_i}\!\bigl(u^2(g)\bigr)
 ~=~\rhoR{_i}\!(a)\,\rhoR{_i}\!(g)\,\rhoR{_i}\!(a)^\dagger
 ~\stackrel{!}{=}~
 \UU[_i]{\rep{r}}\,\UU[_i]{\rep{r}}^*\,\rhoR{_i}\!(g)\,\UU[_i]{\rep{r}}^T\,\UU[_i]{\rep{r}}^\dagger~=~
 V_{\rep[_i]{r}}\,\rhoR{_i}\!(g)\,V_{\rep[_i]{r}}^\dagger\;,
\end{equation}
where
\begin{equation}\label{eq:Vi}
 V_{\rep[_i]{r}}~=~\UU[_i]{\rep{r}}\,\UU[_i]{\rep{r}}^*\;.
\end{equation}
By Schur's lemma,
\begin{equation}
 \rhoR{_i}\!(a)~=~\mathrm{e}^{-\I\,\alpha_i}\,V_{\rep[_i]{r}}
\end{equation}
with some real phase $\alpha_i$. Consider now
\begin{equation}\label{eq:rho(u(a))}
 \rhoR{_i}\!\bigl(u(a)\bigr)
 ~=~
 \UU[_i]{\rep{r}}\,\left(\mathrm{e}^{-\I\,\alpha_i}\,\UU[_i]{\rep{r}}\,\UU[_i]{\rep{r}}^*\right)^*\,\UU[_i]{\rep{r}}^\dagger
 ~=~\mathrm{e}^{\I\,\alpha_i}\,\UU[_i]{\rep{r}}\,\UU[_i]{\rep{r}}^*
 ~=~\mathrm{e}^{2\I\,\alpha_i}\,\rhoR{_i}\!(a)\;.
\end{equation}
Since $u$ is an automorphism,  $u(a)$ and, hence, also
$\mathrm{e}^{2\I\,\alpha_i}\,\mathbbm{1}$ are group elements, i.e.
\begin{equation}
 \exists~c\in\DiscreteGroup~:~\rhoR{_i}\!(c)
 ~=~\mathrm{e}^{2\I\,\alpha_i}\,\mathbbm{1}
 \;.
\end{equation}
Furthermore, it is evident that $c$ is in the center of \DiscreteGroup. This shows that
$a$ is a fixed point of $u$ up to multiplication by an element $c$ of the center,
\begin{equation}
  u(a)~=~c\,a\quad\text{with}~ c \in Z(\DiscreteGroup)\;.
\end{equation}

Furthermore, one can show that if the order of $u$ is even,
$\ord u=2n$, $c'=a^n$ is in $Z(\DiscreteGroup)$. This can be discerned
by repeatedly applying \Eqref{eq:usquare},
\begin{align}
  u^{2n}(g) & ~=~ a^n\, g \, a^{-n} \stackrel{!}{~=~} g \quad\forall~g\in \DiscreteGroup\;.
\end{align}

Using the results obtained so far, one can show that for the two cases of odd
$n$ and of odd--order groups all higher--order class--inverting automorphisms
that square to an inner automorphism are related to an involutory
automorphism in the way specified in \Eqref{eq:innerRelation}.

Let $n=2m+1$, i.e.\ $\ord u=4m+2$. Since $a$ is a fixed point of $u$ up
to an element $c$ of the center of $\DiscreteGroup$, i.e.\ $u(a)=c\,a$, $b=a^m$
is a solution to \Eqref{eq:centrecondition},
\begin{equation}
  b \, u(b) \, a 
  ~=~ a^m \, (a\,c)^m \, a 
  ~=~ a^{2m+1} \, c^m ~=~ c'\,c^m \in Z(\DiscreteGroup)\;.
\end{equation}

The same argument can be used for odd--order groups independently of the
order of the automorphism because in this case the order of $a$ is odd,
$a^{2m+1}=e$ for some natural number $m$.

Another special case are ambivalent groups because they can also have odd--order
class--inverting automorphisms. Hence, let $\ord u=2n+1$. Using \Eqref{eq:usquare} one can show that
\begin{align}
  u^{2n+1}(g) &~=~ g ~=~ u(a)^n\, u(g)\, u(a)^{-n} 
  ~=~ a^n\, c^n\, u(g)\, a^{-n}\, c^{-n} 
  \notag\\
  &~=~ a^n\, u(g)\, a^{-n}\quad\forall~g\in \DiscreteGroup\;,
\end{align}
i.e.\ $u(g)=a^{-n}\, g\, a^{n}$ and $u$ is an inner automorphism.  Thus,
$u$ is connected to the identity automorphism by conjugation with $b=a^n$,
and, for ambivalent groups, the identity automorphism is class--inverting and
involutory.

In conclusion, we have shown that class--inverting automorphisms of higher order
than two that square to inner automorphisms can always be related to physically
equivalent involutory automorphisms if the order of the original automorphism is
odd, or $4m+2$, or if the order of the group is odd. Using the latter result, we
will show in the next section that there are no automorphisms which can be used
as CP transformations for non--Abelian groups of odd order.

\subsection{No class--inverting automorphism for non--Abelian groups of odd order}
\label{app:proofOddOrder}

Let us now show, using the results obtained above, that non--Abelian groups of
odd order do not admit any class--inverting automorphisms that square to inner
automorphisms.\footnote{We cannot exclude the possibility that there are
class--inverting automorphisms that square to an outer automorphism. However, we
have not found an example.} A remarkable implication of this is that non--Abelian groups
of odd order do not admit bases with real Clebsch--Gordan coefficients. The
proof follows the lines of \cite{Sharp:1975}.

One can show \cite{Feit:1967} that any class--inverting automorphism of an odd
order non--Abelian group is fixed--point free. This is because the only
conjugacy class of such groups which contains $g$ and $g^{-1}$ at the same time
is the identity class. Thus, any involutory class--inverting automorphism would
be order two and fixed--point free. However, the existence of such an
automorphism contradicts the assumption that the group is non--Abelian. This can
be seen as follows. Consider a group \DiscreteGroup\ and let $u$ be an
order--two, fixed point free automorphism. Then the map
\begin{equation}
  g~\longmapsto~g^{-1} \, u(g)
\end{equation}
is injective because
\begin{equation}
  g^{-1}\, u(g)~=~h^{-1} \,u(h) 
  \quad \Longleftrightarrow \quad 
  h\, g^{-1}~=~u(h\, g^{-1})\;,
\end{equation}
which is impossible as $u$ is fixed point free. An injective map on a finite set
is automatically bijective, and, hence, one can write every element $g\in
\DiscreteGroup$ as $h^{-1} u(h)$ for some $h\in \DiscreteGroup$. This implies
that the automorphism $u$ acts on group elements as inversion,
\begin{equation}
  u(g) ~=~u(h^{-1}\, u(h))~=~u(h)^{-1} \, h~=~g^{-1}\;.
\end{equation}
However, a group for which inversion is an automorphism can be shown to be
Abelian,
\begin{align}
  g \, h & ~=~ (h^{-1} \,g^{-1})^{-1} 
  ~=~ u(h^{-1}\, g^{-1}) 
  ~=~ u(h^{-1})\,u(g^{-1}) ~=~ h \, g \quad \forall\,g,h \in \DiscreteGroup\;,
\end{align}
which contradicts the assumption that the group is non--Abelian.

Hence, there can be no involutory class--inverting automorphism for
non--Abelian groups of odd order. This immediately implies that there is no
basis with real Clebsch--Gordan coefficients for such groups. 

This result can be extended in the following way. Let $u$ be a
class--inverting automorphism of order greater than two that squares to an inner
automorphism,
\begin{equation}\label{eq:usquare2}
  \exists ~a \in \DiscreteGroup~:~u^2(g) 
  ~=~ 
  a \, g \, a^{-1} \quad \forall~ g \in \DiscreteGroup\;.
\end{equation}
Since the order of \DiscreteGroup\ is odd, there is a natural number $m$ such
that $a^{2m+1}=e$. Then, as shown in the preceding section, the automorphism
\begin{equation}
  u'(g) ~=~ a^m \, u(g) \, a^{-m} \quad \forall~ g \in \DiscreteGroup
\end{equation}
is class--inverting and involutory. However, this leads to a contradiction
because a non--Abelian group of odd order does not possess such an automorphism.
Therefore, no higher--order class--inverting automorphism with the property
\eqref{eq:usquare2} exists.

In summary, odd--order non--Abelian groups do not have a basis with real
Clebsch--Gordan coefficients and do not allow for consistent CP
transformations in generic settings with the possible caveat of automorphisms
that square to outer automorphisms, for which we have found no example, though.

\subsection{The extended twisted Frobenius--Schur indicator}
\label{app:extendedFS}

Here, we state the proof that the extended twisted Frobenius--Schur indicator
\eqref{eq:FSIn} can be used to check whether an automorphism $u$ of arbitrary
order is class--inverting or not. Let $n=\ord{(u)}/2$ for even--order and
$n=\ord{(u)}$ for odd--order automorphisms. Then one can rewrite the
$n^\mathrm{th}$ extended twisted Frobenius--Schur indicator in component
form,
\begin{align}
  \FSI^{(n)}(\rep[_i]{r}) 
  & ~=~ \frac{(\dim{\rep[_i]{r}})^{n-1}}{|\DiscreteGroup|^n}\,
  \sum_{g_1,\dots,g_n \in \DiscreteGroup} \, \chiR{_i}(g_1 \, u(g_1)\cdots g_n \, u(g_n))
  \notag\\
  & ~=~ \frac{(\dim{\rep[_i]{r}})^{n-1}}{|\DiscreteGroup|^n}\,
  \sum_{g_1,\dots,g_n \in \DiscreteGroup} \, \tr\left[\rhoR{_i}(g_1) \, \rhoR{_i}(u(g_1))\cdots \rhoR{_i}(g_n) \, \rhoR{_i}(u(g_n))\right]
  \notag\\
  & ~=~ \frac{(\dim{\rep[_i]{r}})^{n-1}}{|\DiscreteGroup|^n}\,
  \sum_{g_1,\dots,g_n \in \DiscreteGroup} \, \left[\rhoR{_i}(g_1)\right]_{\alpha_1 \beta_1} \, \left[\rhoR{_i}(u(g_1))\right]_{\beta_1 \gamma_1} \delta_{\gamma_1 \alpha_2} \cdots
  \notag\\
  & \hphantom{ ~=~ \frac{(\dim{\rep[_i]{r}})^{n-1}}{|\DiscreteGroup|^n}\,\sum_{g_1,\dots,g_n \in \DiscreteGroup} \, }\qquad\cdots\left[\rhoR{_i}(g_n)\right]_{\alpha_n \beta_n} \, \left[\rhoR{_i}(u(g_n))\right]_{\beta_n \gamma_n}\delta_{\gamma_n\alpha_1}
  \label{eq:FSIncomponents}\;.
\end{align}
By the Schur orthogonality relation \eqref{eq:Schurorthogonality}, this
expression is $0$ if $\rhoR{_i}\!(g)$ and $[\rhoR{_i}\!(u(g))]^*$ are not in
equivalent representations, which is the case for at least one \rep[_i]{r} if
$u$ is not class--inverting.

Assume now that $u$ is class--inverting such that there is a unitary matrix $\UU[_i]{\rep{r}}$ for each
irreducible representation \rep[_i]{r} with
\begin{equation}\label{eq:conjugateEquivalence2}
  \rhoR{_i}\!(u(g)) ~=~ \UU[_i]{\rep{r}} \, \rhoR{_i}\!(g)^* \, \UU[_i]{\rep{r}}^\dagger\;, \qquad \forall ~ i \;.
\end{equation}
This can be used together with the Schur orthogonality relation \eqref{eq:Schurorthogonality} to simplify each
of the factors of the product in \Eqref{eq:FSIncomponents},
\begin{equation}
  \sum_{g \in \DiscreteGroup} \, \left[\rhoR{_i}(g)\right]_{\alpha \beta} \, 
  \left[\rhoR{_i}(u(g))\right]_{\beta \gamma} ~=~ \frac{|\DiscreteGroup|}{\dim{\rep[_i]{r}}}\,\left[\UU[_i]{\rep{r}}^*\right]_{\gamma \beta}\,
  \left[\UU[_i]{\rep{r}}\right]_{\beta \alpha}\;.
\end{equation}
Hence, the extended twisted Frobenius--Schur indicator yields
\begin{align}
  \FSI^{(n)}(\rep[_i]{r}) & ~=~ \frac{1}{\dim{\rep[_i]{r}}}\,\delta_{\gamma_n\alpha_1}\,\left[\UU[_i]{\rep{r}}^*\right]_{\gamma_n \beta_n}\,\left[\UU[_i]{\rep{r}}\right]_{\beta_n \alpha_n}
  \cdots\delta_{\gamma_1 \alpha_2}\,\left[\UU[_i]{\rep{r}}^*\right]_{\gamma_1 \beta_1}\,\left[\UU[_i]{\rep{r}}\right]_{\beta_1 \alpha_1}
  \notag\\
  & ~=~ \frac{1}{\dim{\rep[_i]{r}}}\,\tr{\left[(\UU[_i]{\rep{r}}^* \, \UU[_i]{\rep{r}})^n\right]} ~=~ \frac{1}{\dim{\rep[_i]{r}}}\,\tr{\left[(V_{\rep[_i]{r}})^n\right]}\label{eq:FSInResult}\,.
\end{align}
Due to the cyclicity of the trace it is clear that
$\tr{\left[(\UU[_i]{\rep{r}}^* \,
\UU[_i]{\rep{r}})^n\right]}=\tr{\left[(V_{\rep[_i]{r}})^n\right]}$ is real.
Moreover, inserting \Eqref{eq:conjugateEquivalence2} $2n$ times into itself,
\begin{equation}
  \rhoR{_i}\!(u^{2n}(g)) ~=~ V_{\rep[_i]{r}}^n \, \rhoR{_i}\!(g) \, (V_{\rep[_i]{r}}^n)^\dagger ~=~ \rhoR{_i}\!(g)\;, \qquad \forall ~ i \;,
\end{equation}
where in the last step $u^{2n}=\mathrm{id}$ has been used, one can see that
Schur's lemma implies $V_{\rep[_i]{r}}^n \propto \mathbbm{1}$.  In fact, since
the trace of $V_{\rep[_i]{r}}^n$ is real, the proportionality factor can only be
$\pm 1$.  Plugging this back into \Eqref{eq:FSInResult} completes the proof that
the $n^\mathrm{th}$ extended twisted Frobenius--Schur indicator is $\pm 1$
for all irreducible representations of \DiscreteGroup\  if $u$ is
class--inverting and $0$ for at least one irrep if not.

\bibliography{Orbifold}

\providecommand{\bysame}{\leavevmode\hbox to3em{\hrulefill}\thinspace}
\frenchspacing
\newcommand{\origttfamily}{}
\let\origttfamily=\ttfamily
\renewcommand{\ttfamily}{\origttfamily \hyphenchar\font=`\-}

\begin{thebibliography}{10}

\bibitem{Sakharov:1967dj}
A.~Sakharov, Pisma Zh.Eksp.Teor.Fiz. \textbf{5} (1967), 32.

\bibitem{Kobayashi:1973fv}
M.~Kobayashi and T.~Maskawa, Prog.Theor.Phys. \textbf{49} (1973), 652.

\bibitem{Chen:2009gf}
M.-C. Chen and K.~Mahanthappa, Phys.Lett. \textbf{B681} (2009), 444,
  \texttt{arXiv:0904.1721} [hep-ph].

\bibitem{Bernabeu:1986fc}
J.~Bernabeu, G.~Branco, and M.~Gronau, Phys.Lett. \textbf{B169} (1986), 243.

\bibitem{Gronau:1986xb}
M.~Gronau, A.~Kfir, and R.~Loewy, Phys.Rev.Lett. \textbf{56} (1986), 1538.

\bibitem{Branco:1986gr}
G.~Branco, L.~Lavoura, and M.~Rebelo, Phys.Lett. \textbf{B180} (1986), 264.

\bibitem{Lebedev:2002wq}
O.~Lebedev, Phys.Rev. \textbf{D67} (2003), 015013,
  \texttt{arXiv:hep-ph/0209023} [hep-ph].

\bibitem{Ecker:1981wv}
G.~Ecker, W.~Grimus, and W.~Konetschny, Nucl.Phys. \textbf{B191} (1981), 465.

\bibitem{Ecker:1983hz}
G.~Ecker, W.~Grimus, and H.~Neufeld, Nucl.Phys. \textbf{B247} (1984), 70.

\bibitem{Branco:1999fs}
G.~C. Branco, L.~Lavoura, and J.~P. Silva, Int.Ser.Monogr.Phys. \textbf{103}
  (1999), 1.

\bibitem{Sozzi:2008zza}
M.~S. Sozzi, \emph{{Discrete symmetries and CP violation: From experiment to
  theory}}, 2008.

\bibitem{Holthausen:2012dk}
M.~Holthausen, M.~Lindner, and M.~A. Schmidt, JHEP \textbf{1304} (2013), 122,
  \texttt{arXiv:1211.6953} [hep-ph].

\bibitem{Feruglio:2012cw}
F.~Feruglio, C.~Hagedorn, and R.~Ziegler, JHEP \textbf{1307} (2013), 027,
  \texttt{arXiv:1211.5560} [hep-ph].

\bibitem{Nishi:2013jqa}
C.~Nishi, Phys.Rev. \textbf{D88} (2013), 033010, \texttt{arXiv:1306.0877}
  [hep-ph].

\bibitem{GAP4}
The GAP~Group, \emph{{GAP -- Groups, Algorithms, and Programming, Version
  4.5.5}}, 2012.

\bibitem{Bickerstaff:1985jc}
R.~Bickerstaff and T.~Damhus, International Journal of Quantum Chemistry
  \textbf{XXVII} (1985), 381.

\bibitem{Ramond:2010zz}
P.~Ramond, \emph{Group theory: A physicist's survey}, 2010.

\bibitem{Kawanaka1990}
N.~Kawanaka and H.~Matsuyama, Hokkaido Math.J. \textbf{19} (1990), 495.

\bibitem{Sharp:1975}
W.~Sharp, L.~Biedenharn, E.~de~Vries, and A.~van Zanten, Canad. J. Math.
  \textbf{27} (1974), 246.

\bibitem{Damhus:1981}
T.~Damhus, Journal of Mathematical Physics \textbf{22} (1981), no.~1, 7.

\bibitem{Ecker:1987qp}
G.~Ecker, W.~Grimus, and H.~Neufeld, J.Phys. \textbf{A20} (1987), L807.

\bibitem{Feit:1967}
W.~Feit, \emph{{Characters of finite groups}}, Benjamin, 1967.

\bibitem{Haber:2012np}
H.~E. Haber and Z.~Surujon, Phys.Rev. \textbf{D86} (2012), 075007,
  \texttt{arXiv:1201.1730} [hep-ph].

\bibitem{Branco:1983tn}
G.~Branco, J.~Gerard, and W.~Grimus, Phys.Lett. \textbf{B136} (1984), 383.

\bibitem{Merle:2011vy}
A.~Merle and R.~Zwicky, JHEP \textbf{1202} (2012), 128,
  \texttt{arXiv:1110.4891} [hep-ph].

\bibitem{Adulpravitchai:2009kd}
A.~Adulpravitchai, A.~Blum, and M.~Lindner, JHEP \textbf{0909} (2009), 018,
  \texttt{arXiv:0907.2332} [hep-ph].

\bibitem{Kobayashi:2006wq}
T.~Kobayashi, H.~P. Nilles, F.~Pl{\"o}ger, S.~Raby, and M.~Ratz, Nucl. Phys.
  \textbf{B768} (2007), 135, \texttt{hep-ph/0611020}.

\bibitem{Fischer:2012qj}
M.~Fischer, M.~Ratz, J.~Torrado, and P.~K. Vaudrevange, JHEP \textbf{1301}
  (2013), 084, \texttt{arXiv:1209.3906} [hep-th].

\bibitem{Fischer:2013qza}
M.~Fischer, S.~Ramos-Sanchez, and P.~K.~S. Vaudrevange, JHEP \textbf{1307}
  (2013), 080, \texttt{arXiv:1304.7742} [hep-th].

\bibitem{Berasaluce-Gonzalez:2013bba}
M.~Berasaluce-Gonz{\'a}lez, G.~Ram{\'i}rez, and A.~M. Uranga,
  \texttt{arXiv:1310.5582} [hep-th].

\bibitem{Feruglio:2007uu}
F.~Feruglio, C.~Hagedorn, Y.~Lin, and L.~Merlo, Nucl.Phys. \textbf{B775}
  (2007), 120, \texttt{arXiv:hep-ph/0702194} [hep-ph].

\bibitem{Ma:2001dn}
E.~Ma and G.~Rajasekaran, Phys.Rev. \textbf{D64} (2001), 113012,
  \texttt{arXiv:hep-ph/0106291} [hep-ph].

\bibitem{Ishimori:2010au}
H.~Ishimori, T.~Kobayashi, H.~Ohki, Y.~Shimizu, H.~Okada, et~al.,
  Prog.Theor.Phys.Suppl. \textbf{183} (2010), 1, \texttt{arXiv:1003.3552}
  [hep-th].

\bibitem{Holthausen:2011vd}
M.~Holthausen and M.~A. Schmidt, JHEP \textbf{1201} (2012), 126,
  \texttt{arXiv:1111.1730} [hep-ph].

\bibitem{Ma:2006ip}
E.~Ma, Mod.Phys.Lett. \textbf{A21} (2006), 1917, \texttt{arXiv:hep-ph/0607056}
  [hep-ph].

\end{thebibliography}
\addcontentsline{toc}{section}{Bibliography}
\bibliographystyle{NewArXiv} 
\end{document}